\documentclass[prl,12pt]{revtex4}
\usepackage{amssymb}
\usepackage{amsmath}
\usepackage{graphicx}
\usepackage{color}
\DeclareGraphicsExtensions{.jpg,.jpeg,.pdf,.png,.mps,.eps,.ps}

\newcommand{\beq}{\begin{equation}}
\newcommand{\eeq}{\end{equation}}
\newcommand{\bea}{\begin{eqnarray}}
\newcommand{\ena}{\end{eqnarray}}
\newcommand{\etal}{{\it et al.}}
\newcommand{\ie}{{\it i.e.}}
\newcommand{\eg}{{\it e.g.}}
\newcommand{\etc}{{\it etc.}}
\newcommand{\lsim}{\mathrel{\mathop{\kern 0pt \rlap
{\raise.2ex\hbox{$<$}}}
\lower.9ex\hbox{\kern-.190em $\sim$}}}
\newcommand{\gsim}{\mathrel{\mathop{\kern 0pt \rlap
{\raise.2ex\hbox{$>$}}}
\lower.9ex\hbox{\kern-.190em $\sim$}}}

\newcommand{\npb}[3]{Nucl.\ Phys.\ B\ {\bf #1}, #3 (#2)}

\newcommand{\pr}[3]{Phys.\ Rev.\ {\bf #1}, #3 (#2)}
\renewcommand{\prl}[3]{Phys.\ Rev.\ Lett. {\bf #1}, #3 (#2)}
\renewcommand{\prd}[3]{Phys.\ Rev.\ D\ {\bf #1}, #3 (#2)}

\renewcommand{\pra}[3]{Phys.\ Rev.\ A\ {\bf #1}, #3 (#2)}
\renewcommand{\pl}[3]{Phys.\ Lett.\ {\bf #1}, #3 (#2)}

\newcommand{\href}[2]{#1}

\definecolor{cyan}{cmyk}{1.,0.,0.,0.5}
\definecolor{magenta}{cmyk}{0.,1.,0.,0.5}
\definecolor{verdatre}{cmyk}{0.5,0.,0.5,0.5}
\definecolor{yellow}{cmyk}{0.,0.,0.2,0.0}
\definecolor{rouge}{cmyk}{0.,0.4,0.6,0.0}
\definecolor{orange}{cmyk}{0.,0.5,0.5,0.}
\definecolor{violet}{rgb}{0.5,0.,0.5}

\begin{document}

\noindent
\title{Nambu--Jona-Lasinio Model Revisited with a Simple Regularization-Renormalization Method}
 \vskip 1.cm
\author{Guang-jiong Ni $^{\rm 1,2}$}
\email{\ pdx01018@pdx.edu}
\author{Ji-Feng Yang $^{\rm 3}$}
\email{\ jfyang@phy.ecnu.edu.cn}
\author{Jianjun Xu $^{\rm 2}$}
\email{\ xujj@fudan.edu.cn}
\author{Senyue Lou $^{\rm 4,5}$}
\email{\ lousenyue@nbu.edu.cn}

\affiliation{\vskip 0.25cm $^{\rm 1}$ Department of Physics, Portland State University, Portland, OR97207, U. S. A.\\
$^{\rm 2}$ Department of Physics, Fudan University, Shanghai, 200433, China\\
$^{\rm 3}$ Department of Physics, East China Normal University, Shanghai, 200062, China\\
$^{\rm 4}$ Department of Physics, Ningbo University, Ningbo, 315211, China\\
$^{\rm 5}$ Shanghai Key Laboratory of Trustworthy Computing, East China Normal University, Shanghai, 200062, China}

\vskip 5mm
\date{\today}

\vskip 0.5cm
\begin{abstract}

\noindent
According to the pioneering model proposed by Nambu and Jona-Lasinio (NJL) \cite{2,3}, a massless fermion acquires its mass via a vacuum phase transition (VPT) process. Our discussion is considerably simplified because an effective Hamiltonian for VPT is proposed and a simple regularization-renormalization method (RRM) is adopted. An unambiguous constraint is found as $\frac{\pi^2}{G\Delta^2_1}=\frac{2}{3}$, where $G$ is the coupling constant first introduced in the NJL model while $\Delta_1$ the mass of fermion ($f$) created after the VPT. The masses of bosons with spin-parity $J^P=0^+, 0^-,1^+$ and $1^-$ as collective modes composed of fermion-antifermion pairs ($f\bar{f}$s) are also calculated by the method of random phase approximation (RPA).

{\bf Keywords}:\; Nambu--Jona-Lasinio Model, Vacuum Phase Transition, Regularization-Renormalization-Method, Goldstone Boson\\
{\bf PACS}:\;03.70.+k; 11.10.Gh; 11.30.Qc; 14.80.Bn

\end{abstract}

\maketitle \vskip 1cm

\section{I. Introduction}
\label{sec:introduction}
\setcounter{equation}{0}
\renewcommand{\theequation}{1.\arabic{equation}}

Based on an analogy with the theory for superconductivity by Bardeen, Cooper and Schrieffer (BCS) in 1957 \cite{1}, in 1961, Nambu and Jona-Lasinio (NJL) proposed the first successful dynamical model in particle physics for the mass creation mechanism of an elementary fermion (f) and their collective modes (composed of $f\bar{f}$ pairs) \cite{2,3}. The development of NJL model and its remarkable applications to physics for hadrons and nuclei have been well summarized, to our knowledge, in two review articles \cite{4,5}. However, it seems that two difficulties still remain unsolved. First of them can be seen from the Lagrangian density of NJL model in (3+1) space-time (Bjorken-Drell metric is used with $\hbar=c=1$):
\begin{equation}\label{1-1}
{\cal L}=i\bar{\psi}\gamma^\mu\partial_\mu\psi+G[(\bar{\psi}\psi)^2-(\bar{\psi}\gamma_5\psi)^2]
\end{equation}
where the self-coupling constant $G$ of massless fermion field $\psi$ has a dimension $[mass]^{-2}\sim [M]^{-2}$. So in the conventional quantum field theory (QFT), the perturbative treatment becomes nonrenormalizable. Usually, one has to introduce a cutoff $\Lambda$ in the momentum integration for virtual massless particles. Interestingly, as pointed out by NJL, in order to fit the observed pion coupling constant, the value of $\Lambda$ has to be rather small, being of the same order as the nucleon mass \cite{2}.

Second, NJL model seems difficult to deal with the property of quark confinement when it is used as a phenomenological theory for QCD.

In this paper, we will focus on the first difficulty which bothered one of us (Ni) a lot in a paper \cite{6}, where a restricted range of values for $G\Lambda^2$ is found as:
\begin{equation}\label{1-2}
1>\dfrac{\pi^2}{G\Lambda^2}>[\sqrt2-\ln(1+\sqrt2)]=0.53284
\end{equation}
before the NJL model can be meaningful in (3+1) space-time. Such constraint on the coupling constant $G$ was named as "fine-tuning" by some authors. The ambiguity of Eq.(1.2) lies in the fact that $G$ appears in a product $G\Lambda^2$ with $\Lambda$ being unfixed, showing that as long as we didn't have a satisfied regularization-renormalization method (RRM) for QFT, the ambiguity remains inevitably.

In 1994-1995, one of us (Yang) proposed a simple RRM \cite{7,8,9}, which has been used in various cases in QFT since then, especially in the recent calculation for Higgs mass in the standard model \cite{10,11} with dynamical symmetry breaking (\ie, $\sigma=0$, \cite{12} and references therein). Hence the aim of this paper is to get rid of the cutoff $\Lambda$ by using our simple RRM, arriving at an unambiguous constraint that
\begin{equation}\label{1-3}
\dfrac{\pi^2}{G\Delta_1^2}=\dfrac{2}{3}
\end{equation}
where $\Delta_1$ is the observable mass of fermion created after the vacuum phase transition (VPT) in the NJL model, Eq.(1.1).

The organization of this paper is as follows: In section II, an effective Hamiltonian for VPT in NJL model is proposed. Section III is basically a review of NJL vacuum state (NJLVS) and its formal solution. How the result Eq.(1.3) will be found by a simple RRM is discussed in section IV. Section V is devoted to discuss why the NJL transformation (NJLT) initiated also by NJL can be developed into a systematic method. Then the masses of scalar and vector bosons as collective modes after VPT are evaluated in NJL model in sections VI and VII. The final section VIII contains the summary and discussions. Some details are added at three Appendices.

\section{II. Effective Hamiltonian for Vacuum Phase Transition in NJL Model}
\label{sec:hamiltonian}
\setcounter{equation}{0}
\renewcommand{\theequation}{2.\arabic{equation}}

As in Ref.\cite{6} (see also \cite{13}), we first promote the massless fermion field $\psi$ in Eq.(1.1) into field operator $\hat{\psi}$ at the QFT level (in Heisenberg picture):
\begin{equation}\label{2-1}
\hat{\psi}({\bf x},t)=\frac{1}{\sqrt{V}}\sum_{{\bf p},h}[\hat{a}_{{\bf p},h}(t)u_{{\bf p},h}e^{i{\bf p}\cdot{\bf x}}
+\hat{b}^\dag_{{\bf p},h}(t)v_{{\bf p},h}e^{-i{\bf p}\cdot{\bf x}}]
\end{equation}
where $\hat{a}_{{\bf p},h}(t)\;(\hat{b}^\dag_{{\bf p},h}(t))$ is the annihilation (creation) operator of particle (antiparticle) in the Fock space with ${\bf p}$ and $h(=\pm1)$ being momentum and helicity respectively, satisfying the anticommutation relations as:
\begin{equation}\label{2-2}
[\hat{a}_{{\bf p},h}(t),\hat{a}^\dag_{{\bf p'},h'}(t)]_+=[\hat{b}_{{\bf p},h}(t),\hat{b}^\dag_{{\bf p'},h'}(t)]_+=\delta_{\bf pp'}\delta_{hh'}
\end{equation}
\etc ~Then we get the Hamiltonian operator as (\cite{6,13})
\begin{equation}\label{2-3}
H=\int_V{\cal H}d^3x\to :\hat{H}(\hat{a}_{{\bf p},h},\hat{a}^\dag_{{\bf p'},h'},\hat{b}_{{\bf p},h},\hat{b}^\dag_{{\bf p'},h'},\cdots):
\end{equation}
where :\ : means the normal product ordering is made.

Now in relativistic QFT, the method of "motion equation of the Green function (MEGF)" which works so effectively in nonrelativistic superconductivity theory (\cite{14,15,16}, see also Appendix 8A in Ref.\cite{17}), can be used to calculate the expectation value of Eq.(2.3) in the physical vacuum $|\tilde{0}\rangle$, the relativistic counterpart of the superconductivity ground state, yielding the vacuum energy $E_{vac}$ as:
\begin{equation}\label{2-4}
\langle\tilde{0}|:\hat{H}:|\tilde{0}\rangle=E_{vac}=4\sum_{\bf p}|{\bf p}|n({\bf p})-\dfrac{16G}{V}\left[\sum_{\bf p}\nu({\bf p})\right]^2
\end{equation}
Here
\begin{equation}\label{2-5}
n({\bf p})=\langle\tilde{0}|\hat{a}^\dag_{{\bf p},h}\hat{a}_{{\bf p},h}|\tilde{0}\rangle=\langle\tilde{0}|\hat{b}^\dag_{{\bf p},h}\hat{b}_{{\bf p},h}|\tilde{0}\rangle \qquad ({\bf p},h \, \text{not summed})
\end{equation}
is the condensation number of massless fermion ($f$) or antifermion ($\bar{f}$) in the vacuum (with volumn $V$), whereas \footnotemark[1]\footnotetext[1]{See Eq.(3.7) below for a rigorous definition.}
\begin{equation}\label{2-6}
\nu({\bf p})=\langle\tilde{0}|\hat{a}^\dag_{{\bf p},h}\hat{b}^\dag_{{-\bf p},h}|\tilde{0}\rangle=\langle\tilde{0}|\hat{b}_{{-\bf p},h}\hat{a}_{{\bf p},h}|\tilde{0}\rangle
\end{equation}
is the creation or annihilation number of $f\bar{f}$ pairs with zero momentum and angular-momentum in the vacuum. In deriving Eq.(2.4), a "pairing cutoff approximation" (PCA) has been used. The physical implication of Eq.(2.4) prompts us to propose an effective Hamiltonian for vacuum phase transition (VPT) in NJL model as follows:
\begin{eqnarray}
  \hat{H} &=&=\int{\hat{\cal H}}d^3x =\hat{H}_0+\hat{H}_I\\[3mm]
  \hat{H}_0 &=& \sum_{{\bf p},h}\omega_p(\hat{a}^\dag_{{\bf p}h}\hat{a}_{{\bf p}h}+\hat{b}^\dag_{{\bf p}h}\hat{b}_{{\bf p}h}),\quad (\omega_p=|{\bf p}|) \\[3mm]
  \hat{H}_I&=&-\dfrac{G}{V}\sum_{{\bf p,p'},h,h'}hh':[(\hat{a}^\dag_{{\bf p}h}\hat{b}^\dag_{{-\bf p}h}+\hat{b}_{-{\bf p}h}\hat{a}_{{\bf p}h})(\hat{a}^\dag_{{\bf p'}h'}\hat{b}^\dag_{{-\bf p'}h'}+\hat{b}_{-{\bf p'}h'}\hat{a}_{{\bf p'}h'})]:
\end{eqnarray}
Here, first, the Hamiltonian density $\hat{\cal H}$ in relativistic QFT must be invariant under an operation of "strong reflection", \ie, the "space-time inversion" first invented by Pauli in 1955 \cite{18} and further discussed in Refs.\cite{19,20} in the sense of
\begin{equation}\label{2-10}
\hat{\cal H}({\bf x},t)\to\widehat{\cal PT}\hat{\cal H}({\bf x},t)(\widehat{\cal PT})^{-1}=\hat{\cal H}(-{\bf x},-t)=\hat{\cal H}({\bf x},t)
\end{equation}
where
\begin{eqnarray}
  \widehat{\cal PT}(\hat{a}^\dag_{{\bf p}h}\hat{b}^\dag_{{-\bf p}h})(\widehat{\cal PT})^{-1} &=&
  \widehat{\cal PT}\hat{b}^\dag_{-{\bf p}h}(\widehat{\cal PT})^{-1}\widehat{\cal PT}\hat{a}^\dag_{{\bf p}h}(\widehat{\cal PT})^{-1}
  = \hat{a}_{-{\bf p}-h}\hat{b}_{{\bf p}-h}\\
  \widehat{\cal PT}\hat{b}^\dag_{-{\bf p}h}(\widehat{\cal PT})^{-1} &=&\hat{a}_{-{\bf p}-h},\quad\widehat{\cal PT}\hat{a}^\dag_{{\bf p}h}(\widehat{\cal PT})^{-1}=\hat{b}_{{\bf p}-h}
\end{eqnarray}
(See Refs.\cite{19,20}). Note the reversed helicity and the reversed order of an operator product under the "strong reflection".

Second, the Hamiltonian density is also invariant under an operation of hermitian conjugation (h.c.) (\cite{18,19,20}) as:
\begin{equation}\label{2-13}
\hat{\cal H}({\bf x},t)\to\hat{\cal H}^\dag({\bf x},t)=\hat{\cal H}({\bf x},t)
\end{equation}
Third, besides the inversion in (3+1) dimensional space-time as shown in Eq.(2.10), we should consider the pure space inversion (${\bf x}\to -{\bf x},t\to t$), \ie, the parity symmetry $P$ at the level of relativistic quantum mechanism (RQM) and its counterpart $\hat{U}(P)$ at the level of QFT (see Eq.(4.120) in Ref.\cite{21})
\begin{equation}\label{2-14}
\begin{array}{l}
  \hat{U}(P)\hat{a}^\dag_{{\bf p}h}\hat{U}^{-1}(P)=\hat{a}^\dag_{-{\bf p}-h} \\[3mm]
  \hat{U}(P)\hat{b}^\dag_{-{\bf p}h}\hat{U}^{-1}(P)=-\hat{b}^\dag_{{\bf p}-h}
\end{array}
\end{equation}
where the "-" sign means that the "intrinsic parity" of antifermion is opposite to that of fermion. Thus
\begin{eqnarray}
  \hat{U}(P)h\hat{a}^\dag_{{\bf p}h}\hat{b}^\dag_{-{\bf p}h}\hat{U}^{-1}(P)&=&h\hat{U}(P)\hat{a}^\dag_{{\bf p}h}\hat{U}^{-1}(P)\hat{U}(P)\hat{b}^\dag_{-{\bf p}h}\hat{U}^{-1}(P)=-h\hat{a}^\dag_{-{\bf p}-h}\hat{b}^\dag_{{\bf p}-h} \\
  \hat{U}(P)\hat{\cal H}({\bf x},t)\hat{U}^{-1}(P)&=&\hat{\cal H}(-{\bf x},t)=\hat{\cal H}({\bf x},t)
\end{eqnarray}
Hence, as a whole, the Hamiltonian Eq.(2.7) remains a scalar not only in 4D space-time, but also in 3D space. We will generalize it to some species of fermion in further study. But in this paper we just focus on one species as shown by original NJL model, Eq.(1.1), and devote to solve the problem of VPT starting from Eq.(2.7), which includes three kinds of elementary interactions in the vacuum as shown by Fig.1.

\section{III. Nambu--Jona-Lasinio Vacuum State and Its Formal Solution}
\label{sec:vacuum}
\setcounter{equation}{0}
\renewcommand{\theequation}{3.\arabic{equation}}

In dealing with the VPT, Nambu and Jona-Lasinio (NJL) first proposed that the vacuum state $|\tilde{0}\rangle$ in Eqs.(2.4)-(2.6) can be expressed by (see Eq.(3.17) in Ref.\cite{2}):
\begin{equation}\label{3-1}
|\tilde{0}\rangle=\prod_{{\bf p},h}(U_p+hV_p\hat{a}^\dag_{{\bf p}h}\hat{b}^\dag_{-{\bf p}h})|0\rangle
\end{equation}
where the naive vacuum $|0\rangle$ is defined by
\begin{equation}\label{3-2}
\hat{a}_{{\bf p}h}|0\rangle=\hat{b}_{{\bf p}h}|0\rangle=0
\end{equation}
while $U_p$ and $V_p$ are unknown functions of momentum $\bf p$ but assumed to be independent of its direction and the value of $h$. To our knowledge, the helicity $h$ in front of $V_p$ was first added in Ref.\cite{22} but it was missed in our previous paper \cite{23}. We will name Eq.(3.1) as the NJL vacuum state (NJLVS), which is the relativistic counterpart of the BCS ground state for superconductivity (\cite{1}, see also Eq.(8.5.38) in Ref.\cite{17}). Notice that $\langle\tilde{0}|\equiv(|\tilde{0}\rangle)^\dag$, but also we have
\begin{equation}\label{3-3}
\widehat{\cal PT}|\tilde{0}\rangle=\langle\tilde{0}|
\end{equation}
where Eq.(2.11) and the arbitrary definitions of dummy indices ${\bf p}$ and $h$ have been used. Moreover, the existence of $h$ in front of $V_p$ in Eq.(3.1) ensures that
\begin{equation}\label{3-4}
\hat{U}(P)|\tilde{0}\rangle=|\tilde{0}\rangle
\end{equation}
Then the normalization of NJLVS $\langle\tilde{0}|\tilde{0}\rangle=1$ leads to
\begin{equation}\label{3-5}
U_p^2+V_p^2=1
\end{equation}
where Eqs.(2.2) and (3.2) have been used. Now it is easy to prove that
\begin{equation}\label{3-6}
\langle\tilde{0}|\hat{a}^\dag_{{\bf p}h}\hat{a}_{{\bf p}h}|\tilde{0}\rangle=\langle\tilde{0}|\hat{b}^\dag_{{\bf p}h}\hat{b}_{{\bf p}h}|\tilde{0}\rangle=V_p^2=n({\bf p})
\end{equation}
coinciding with Eq.(2.5) derived from the MEGF used for NJL model \cite{6}. However, the "condensation number of $f\bar{f}$ pairs in the NJLVS" should be rigorously expressed as
\begin{equation}\label{3-7}
\nu({\bf p})=\langle\tilde{0}|h\hat{a}^\dag_{{\bf p}h}\hat{b}^\dag_{-{\bf p}h}|\tilde{0}\rangle=\langle\tilde{0}|h\hat{b}_{-{\bf p}h}\hat{a}_{{\bf p}h}|\tilde{0}\rangle=U_pV_p
\end{equation}
such that Eq.(3.7) is an invariant number under the "strong reflection" due to Eq.(2.11).

Furthermore, we can calculate the vacuum energy from E.(2.7). It is easy to obtain
\begin{equation}\label{3-8}
\langle\tilde{0}|\hat{H}_0|\tilde{0}\rangle=4\sum_{\bf p}\omega_pV_p^2
\end{equation}
The coefficient 4 comes from the sum over $f$ and $\bar{f}$ each with helicity $h=1$ and $h=-1$. But the calculation on $\hat{H}_I$, Eq.(2.9), needs to be careful. After normal ordering, we have
\begin{eqnarray}
  \hat{H}_I &=&{\hat{H}_I}'+{\hat{H}_I}'' \\[3mm]
  &&\begin{array}{l}\hspace*{-14mm}{\hat{H}_I}'= -\dfrac{G}{V}\sum\limits_{\substack{{\bf p,p'}\\ h,h'}}                                                            (hh')[\hat{a}^\dag_{{\bf p}h}\hat{b}^\dag_{-{\bf p}h}\hat{b}_{-{\bf p'}h'}\hat{a}_{{\bf p'}h'}+\hat{a}^\dag_{{\bf p'}h'}\hat{b}^\dag_{-{\bf p'}h'}\hat{b}_{-{\bf p}h}\hat{a}_{{\bf p}h}\\[3mm]
  \qquad\quad+\hat{a}^\dag_{{\bf p}h}\hat{b}^\dag_{-{\bf p}h}\hat{a}^\dag_{{\bf p'}h'}\hat{b}^\dag_{-{\bf p'}h'}
   +\hat{b}_{-{\bf p}h}\hat{a}_{-{\bf p}h}\hat{b}_{-{\bf p'}h'}\hat{a}_{{\bf p'}h'}]\end{array} \\[3mm]
  {\hat{H}_I}'' &=& \dfrac{G}{V}\sum_{\substack{{\bf p,p'}\\ h,h'}}(hh')\delta_{\bf pp'}\delta_{hh'}(\hat{a}^\dag_{{\bf p}h}\hat{a}_{{\bf p}h}+\hat{b}^\dag_{{\bf p'}h'}\hat{b}_{{\bf p'}h'})
\end{eqnarray}
For calculating on ${\hat{H}_I}'$, two cases, either (${\bf p}={\bf p'},h=h'$) or (${\bf p}\neq{\bf p'},\text{and/or}\,\,h\neq h'$) need to be separated. For the former case, we easily get
\begin{equation}\label{3-12}
\langle\tilde{0}|\hat{a}^\dag_{{\bf p}h}\hat{b}^\dag_{-{\bf p}h}\hat{b}_{-{\bf p'}h'}\hat{a}_{{\bf p'}h'}|\tilde{0}\rangle=V_p^2\delta_{\bf p,p'}\delta_{hh'}
\end{equation}
But for the latter case, we have to add a factor $(1-\delta_{\bf pp'}\delta_{hh'})$ by hand and obtain that
\begin{equation}\label{3-13}
\langle\tilde{0}|{\hat{H}_I}'|\tilde{0}\rangle=-\dfrac{G}{V}\left[16\sum_{\bf p,p'}(U_pV_p)(U_{p'}V_{p'})-4\sum_{\bf p}(U_p^2V_p^2
-V_p^4)\right]
\end{equation}
where the last term will be combined with the contribution of
\begin{equation}\label{3-14}
\langle\tilde{0}|{\hat{H}_I}''|\tilde{0}\rangle=\dfrac{4G}{V}\sum_{\bf p}V_p^2
\end{equation}
So eventually, the vacuum energy reads
\begin{equation}\label{3-15}
E_{vac}=\langle\tilde{0}|\hat{H}|\tilde{0}\rangle=4\sum_{\bf p}\omega_pV_p^2-\dfrac{16G}{V}\left[\sum_{\bf p,p'}(U_pV_p)(U_{p'}V_{p'})\right]+\dfrac{8G}{V}\sum_{\bf p}U_p^2V_p^2
\end{equation}
In view of Eqs.(3.6) and (3.7), Eq.(3.15) is exactly the same as Eq.(2.4) we got from the method of MEGF except the extra term
$\frac{8G}{V}\sum_{\bf p}U_p^2V_p^2$. But since $\sum_{\bf p}\to \frac{V}{(2\pi)^3}\int d^3p$, if the fermion, like a lepton, can move in a space $V\to\infty$, \footnotemark[1]\footnotetext[1]{In this paper, we consider leptons $e,\mu$ and $\tau$ with neutrinos excluded.} this term can be safely ignored with respect to the other term containing $G$. We will only keep this term for quark confinement problem where $V$ is finite.

Evidently, if $G\to0, |\tilde{0}\rangle\to |0\rangle, V_p^2=0$ and $E_{vac}=0$. However, once $G>0$, will we have $E_{vac}<0$ to favor a VPT?

For this purpose, we take the partial derivative of Eq.(3.15) with respect to $V_p$
\begin{equation}\label{3-16}
\dfrac{\partial}{\partial V_p}E_{vac}=8\omega_pV_p-\dfrac{32G}{V}\left(\sum_{\bf p}U_pV_p\right)\left(U_p-\dfrac{V_p^2}{U_p}\right)
\end{equation}
by using $\frac{\partial}{\partial V_p}U_p=-\frac{V_p}{U_p}$ due to Eq.(3.5). Thus the condition that $\frac{\partial}{\partial V_p}E_{vac}=0$ leads to
\begin{equation}\label{3-17}
2\omega_pU_pV_p=\dfrac{8G}{V}\left(\sum_{\bf p'}U_{\bf p'}V_{\bf p'}\right)(U_p^2-V_p^2)
\end{equation}
The formal solution to Eq.(3.17) is similar to that in BCS theory for superconductivity (\cite{1}, \cite{16}, see also section 8.5B in \cite{17}). Define
\begin{equation}\label{3-18}
\Delta=\dfrac{8G}{V}\left(\sum_{\bf p}U_{\bf p}V_{\bf p}\right)
\end{equation}
which was called as the "energy gap" in BCS theory and now, as will be seen quickly, is just the observable mass of fermion (say, electron) after the VPT ($\Delta=m_e$). Then according to Eq.(3.5), one may parameterize the $V_p$ and $U_p$ by
\begin{equation}\label{3-19}
U_p=\cos\theta_p,\quad V_p=\sin\theta_p
\end{equation}
So Eq.(3.17) reads
\begin{equation}\label{3-20}
\omega_p\sin2\theta_p=\Delta\cos2\theta_p
\end{equation}
with its solution being
\begin{equation}\label{3-21}
U_p^2=\dfrac{1}{2}\left(1+\dfrac{\omega_p}{E_p}\right),\quad V_p^2=\dfrac{1}{2}\left(1-\dfrac{\omega_p}{E_p}\right),\quad U_p^2-V_p^2=\dfrac{\omega_p}{E_p}
\end{equation}
where
\begin{equation}\label{3-22}
E_p=\sqrt{\omega_p^2+\Delta^2},\quad (2U_pV_p=\dfrac{\Delta}{E_p})
\end{equation}
is just the energy of massive fermion created after the VPT \footnotemark[1]\footnotetext[1]{We will add the subscript "1" to stress the  $\Delta_1$ being the observable mass after VPT. The quasiparticle's energy $E_p$ shows up as a pole in the Green function at \cite{6}. Eq.(3.21) is shown in the Fig.2.}. And Eq.(3.18) becomes
\begin{equation}\label{3-23}
1=\dfrac{4G}{V}\sum_{\bf p}\dfrac{1}{E_p}=\dfrac{2G}{\pi^2}I
\end{equation}
with the integral $I$ being
\begin{equation}\label{3-24}
I=\int_0^\infty dp\dfrac{p^2}{\sqrt{p^2+\Delta^2}}
\end{equation}
and will be named, like that in BCS theory, as the "gap equation". But can we fix the value of $\Delta$ into $\Delta_1$ via Eq.(3.24)? The answer is "no". See next section.

\section{IV. The Regularization of Divergence and Criterions for VPT}
\label{sec:divergence}
\setcounter{equation}{0}
\renewcommand{\theequation}{4.\arabic{equation}}

We are dealing with the VPT as a dynamical process with $\Delta$ being the "running" order parameter which remains a variable until it is fixed as  $\Delta_1$ via a variational method.

Now comes the main difficulty in our paper --- the integral in Eq.(3.24) is a divergent one, so are the integrals appearing in the vacuum energy, Eq.(3.15). We have been learning the QFT for decades and were bothered by divergences a lot until we use the RRM (proposed first by Yang in Refs.\cite{7,8,9}) as follows. Because the integral $I$ is so-called "quadratic divergent", we first take a partial derivative with respect to the parameter $\Delta^2$:
\begin{equation}\label{4-1}
\dfrac{\partial I}{\partial \Delta^2}=-\dfrac{1}{2}\int_0^\infty \dfrac{p^2 dp}{(p^2+\Delta^2)^{3/2}}
\end{equation}
which becomes only "logarithmically divergent". Then take derivative one more time, yielding a convergent integral as
\begin{equation}\label{4-2}
\dfrac{\partial^2 I}{\partial (\Delta^2)^2}=\dfrac{3}{4}\int_0^\infty \dfrac{p^2 dp}{(p^2+\Delta^2)^{5/2}}=\dfrac{1}{4\Delta^2}
\end{equation}
Now we integrate Eq.(4.2) with respect to $\Delta^2$, returning back to
\begin{equation}\label{4-3}
\dfrac{\partial I}{\partial \Delta^2}=\dfrac{1}{4}\left[\ln(\Delta^2)+C_1\right]
\end{equation}
where an arbitrary constant $C_1$ is added according to the rule of the "indefinite integration" we learned when we were freshmen at universities. However, $\Delta^2$ has the dimension of $(mass)^2\sim [M]^2$, so both $\ln(\Delta^2)$ and $C_1$ are ambiguous or meaningless to mathematicians, who can only accept variables as numbers without any dimension. Hence we have no choice but rewrite $C_1$ in an ambiguous way as $C_1=-\ln\mu_s^2$ such that
\begin{equation}\label{4-4}
\dfrac{\partial I}{\partial \Delta^2}=\dfrac{1}{4}\ln\left(\dfrac{\Delta^2}{\mu_s^2}\right)
\end{equation}
can be accepted in mathematics with $\mu_s$ as an arbitrary "mass scale" to be fixed later in physics. So next step is straightforward:
\begin{equation}\label{4-5}
I=\displaystyle\int\dfrac{\partial I}{\partial\Delta^2}d\Delta^2=\dfrac{1}{4}\left[\Delta^2\left(\ln\dfrac{\Delta^2}{\mu_s^2}-1
\right)+C_2\right]
\end{equation}
where $C_2$ is the second arbitrary constant with dimension $[M]^2$. Eq.(4.5) accomplishes the so-called "regularization" procedure for the divergent integral in Eq.(3.24), ending up with two arbitrary constants $\mu_s$ and $C_2$ waiting to be fixed later in physics. Notice that the combination of Eqs.(4.5), (3.23) and (3.24) gives us that
\begin{equation}\label{4-6}
\left[\Delta^2\left(\ln\dfrac{\Delta^2}{\mu_s^2}-1\right)+C_2\right]_{VPT}=\dfrac{2\pi^2}{G},\ C_2=\dfrac{2\pi^2}{G}-\Delta_1^2\left(\ln\dfrac{\Delta_1^2}{\mu_s^2}-1\right)
\end{equation}
However, we will see later that Eq.(4.6) can only be used after the VPT is finished and thus $\Delta_1^2$ created and fixed with certainty, not before.

Furthermore, we examine another divergent integral contained in the "kinetic energy" term in the vacuum energy, Eq.(3.15):
\begin{equation}\label{4-7}
4\sum_{\bf p}|{\bf p}|n({\bf p})=\dfrac{V}{\pi^2}J+\dfrac{\Delta^2V}{\pi^2}I
\end{equation}
where
\begin{equation}\label{4-8}
J=\int_0^\infty dp(p^3-p^2\sqrt{p^2+\Delta^2})
\end{equation}
is so-called "quartic divergent". So it needs to be handled as follows:
\begin{equation}\label{4-9}
\dfrac{\partial^3J}{\partial(\Delta^2)^3}=-\dfrac{1}{8\Delta^2}
\end{equation}
\begin{equation}\label{4-10}
J=-\dfrac{1}{16}\left[\Delta^4\left(\ln\dfrac{\Delta^2}{\mu_s^2}-\dfrac{3}{2}\right)+2C'_2\Delta^2+C_3\right]
\end{equation}
One thing is important here that the $\mu_s^2$ remains the same as that appeared in the integral $I$, Eq.(4.5). This is because as a "mass scale" in one theory like NJL model here, the $\mu_s$ must be unified for all divergent integrals. However, on the other hand, there is no a priori reason for the constant $C'_2$ here being equal to $C_2$ in Eq.(4.5). \footnotemark[1]\footnotetext[1]{If we try another scheme: $4\sum_{\bf p}|{\bf p}|n({\bf p})=\frac{V}{\pi^2}K,\, K=\int_0^\infty(p^3-\frac{p^4}{\sqrt{p^2+\Delta^2}})dp$, we would find a $\tilde{C}_2=\frac{2\pi^2}{3G}-\frac{1}{3}\Delta^2\ln\frac{\Delta^2}{\mu_s^2}+\Delta^2$ in $K$, $\tilde{C}_2\neq C_2$ as shown by Eq.(4.6). So only Eq.(4.10) will be treated below.}
The third arbitrary constant here for NJL model is trivial, because the condition that
\begin{equation}\label{4-11}
E_{vac}(\Delta^2\to 0)=\langle0|\hat{H}|0\rangle=0\to C_3=0
\end{equation}
Hence
\begin{equation}\label{4-12}
\begin{array}{l}
E_{vac}=-\dfrac{V}{16\pi^2}\left\{\Delta^4\left(\ln\dfrac{\Delta^2}{\mu_s^2}-\dfrac{3}{2}\right)+2C'_2\Delta^2-4\Delta^2\left[
\Delta^2\left(\ln\dfrac{\Delta^2}{\mu_s^2}-1\right)+C_2\right]\right.\\[6mm]
+\left.\dfrac{G\Delta^2}{\pi^2}\left[\Delta^2\left(\ln\dfrac{\Delta^2}{\mu_s^2}-1\right)+C_2\right]^2\right\}
\end{array}\end{equation}
should be looked as a function of 4 unknown parameters: $\Delta, \mu_s, C_2, C'_2$ with another constraint gives by the gap equation, Eq.(4.6) once the VPT is achieved. So we need 3 equations derived from Eq.(4.12). The first one seems evident for VPT could occur. Just evaluate that
\begin{equation}\label{4-13}
\begin{array}{l}
\dfrac{\partial E_{vac}}{\partial\Delta^2}=-\dfrac{V}{16\pi^2}\left\{-2\Delta^2\left(\ln\dfrac{\Delta^2}{\mu_s^2}+1\right)+2C'_2-4\left[
\Delta^2\left(\ln\dfrac{\Delta^2}{\mu_s^2}-1\right)+C_2\right]\right.\\[6mm]
+\left.\dfrac{G}{\pi^2}\left[\Delta^2\left(\ln\dfrac{\Delta^2}{\mu_s^2}-1\right)+C_2\right]^2
+\dfrac{2G\Delta^2}{\pi^2}\left[\Delta^2\left(\ln\dfrac{\Delta^2}{\mu_s^2}-1\right)+C_2\right]\ln\dfrac{\Delta^2}{\mu_s^2}\right\}
\end{array}\end{equation}
which is still a function of 4 parameters in the varying process of VPT. The latter is finished only when the gap equation Eq.(4.6) is substituted into Eq.(4.13) and so we demand that
\begin{equation}\label{4-14}
\left.\dfrac{\partial E_{vac}}{\partial\Delta^2}\right|_{VPT}=-\dfrac{V}{8\pi^2}\left[\Delta^2\left(\ln\dfrac{\Delta^2}{\mu_s^2}-1\right)
-\dfrac{2\pi^2}{G}+C'_2\right]=0
\end{equation}
Thus we find, to our surprising pleasure, that
\begin{equation}\label{4-15}
C'_2=C_2=\dfrac{2\pi^2}{G}-\Delta_1^2\left(\ln\dfrac{\Delta_1^2}{\mu_s^2}-1\right)
\end{equation}
(see Eq.(4.6)). In the mean time, the condition that
\begin{equation}\label{4-16}
E_{vac}|_{VPT}=\dfrac{V}{16\pi^2}\Delta_1^4\left(\ln\dfrac{\Delta_1^2}{\mu_s^2}-\dfrac{1}{2}\right)<0
\end{equation}
gives a constraint as
\begin{equation}\label{4-17}
\ln\dfrac{\Delta_1^2}{\mu_s^2}<\dfrac{1}{2}
\end{equation}
Eqs.(4.14)-(4.17) ensure the stability of new physical vacuum after VPT. So there are only two remaining parameters $\Delta_1$ and $\mu_s$ waiting to be fixed under the constraint condition, Eq.(4.17). Let's go ahead, evaluating
\begin{equation}\label{4-18}
\dfrac{\partial^2 E_{vac}}{\partial(\Delta^2)^2}\!=\!\dfrac{V}{16\pi^2}\!\left\{6\ln\dfrac{\Delta^2}{\mu_s^2}\!+\!4\!-\!\dfrac{4G}{\pi^2}\!\left[
\Delta^2\!\left(\ln\dfrac{\Delta^2}{\mu_s^2}\!-\!1\right)\!+\!C_2\right]\!\left(
\ln\dfrac{\Delta^2}{\mu_s^2}\!+\!\dfrac{1}{2}\right)
\!-\!\dfrac{2G\Delta^2}{\pi^2}\!\left(\!\ln\dfrac{\Delta^2}{\mu_s^2}\right)^2\right\}
\end{equation}
Then the substitution of gap equation, Eq.(4.6), into Eq.(4.18) leads to the possibility that
\begin{equation}\label{4-19}
\left.\dfrac{\partial^2 E_{vac}}{\partial(\Delta_1^2)^2}\right|_{VPT}=-\dfrac{V}{8\pi^2}\left(\ln\dfrac{\Delta_1^2}{\mu_s^2}\right)
\left(1+\dfrac{G\Delta_1^2}{\pi^2}\ln\dfrac{\Delta_1^2}{\mu_s^2}\right)\geq0
\end{equation}
If assume $\frac{1}{2}>\ln\frac{\Delta_1^2}{\mu_s^2}>0$ (see Eq.(4.17)), Eq.(4.19) would imply $\left(1+\frac{G\Delta_1^2}{\pi^2}\ln\frac{\Delta_1^2}{\mu_s^2}\right)<0$, or $\ln\frac{\Delta_1^2}{\mu_s^2}<-\frac{\pi^2}{G\Delta_1^2}$, which contradicts the assumption. So we have to admit that
\begin{equation}\label{4-20}
\ln\dfrac{\Delta_1^2}{\mu_s^2}<0
\end{equation}
and $\left(1+\frac{G\Delta_1^2}{\pi^2}\ln\frac{\Delta_1^2}{\mu_s^2}\right)>0$, or
\begin{equation}\label{4-21}
\ln\dfrac{\Delta_1^2}{\mu_s^2}\geq-\dfrac{\pi^2}{G\Delta_1^2}
\end{equation}
As discussed above, we need an equation. Hence we have no choice but set the equal sign only, yielding
\begin{equation}\label{4-22}
\Delta_1^2=\mu_s^2\exp\left(-\dfrac{\pi^2}{G\Delta_1^2}\right)
\end{equation}
which is reasonable since $G>0$ implies the attractive force within a pair of $f\bar{f}$ and when $G\to 0^+,\, \Delta^2\to0$ whereas if $G\to \infty,\, \Delta_1^2\to\mu_s^2$. So it shows clearly that VPT is a nonperturbative process in the sense of $G=0$ being an essential singularity --- no $G\to 0^-$ (repulsive force between $f$ and $\bar{f}$) is allowed. Similar thing happens in the BCS theory for superconductivity (\cite{1}, see \eg, Eq.(8.5.36) in \cite{17}).

As one more equation is needed, we evaluate further that
\begin{equation}\label{4-23}
\dfrac{\partial^3 E_{vac}}{\partial(\Delta^2)^3}=\dfrac{V}{8\pi^2}\left\{\dfrac{3}{\Delta^2}-\dfrac{3G}{\pi^2}\left[
\left(\ln\dfrac{\Delta^2}{\mu_s^2}\right)^2+\ln\dfrac{\Delta^2}{\mu_s^2}\right]-\dfrac{2G}{\pi^2\Delta^2}\left[\Delta^2\left(
\ln\dfrac{\Delta^2}{\mu_s^2}-1\right)+C_2\right]\right\}
\end{equation}
By using Eq.(4.6), the condition for VPT reads
\begin{equation}\label{4-24}
\left.\dfrac{\partial^3 E_{vac}}{\partial(\Delta^2)^3}\right|_{VPT}=-\dfrac{V}{8\pi^2\Delta_1^2}\left\{1+\dfrac{3G\Delta_1^2}{\pi^2}\left[
\left(\ln\dfrac{\Delta_1^2}{\mu_s^2}\right)^2+\ln\dfrac{\Delta_1^2}{\mu_s^2}\right]\right\}\geq0
\end{equation}
or
\begin{equation}\label{4-25}
f(y)=y^2+y+\dfrac{1}{3}a\leq0,\quad (y=\ln\dfrac{\Delta_1^2}{\mu_s^2}<0, a=\dfrac{\pi^2}{G\Delta_1^2}>0)
\end{equation}
The equation $f(y)=0$ has two roots
\begin{equation}\label{4-26}
y_\pm=\dfrac{1}{2}\left[-1\pm\sqrt{1-\dfrac{4a}{3}}\right]<0
\end{equation}
The condition for $y_\pm$ being real imposes a constraint on the parameter $a$
\begin{equation}\label{4-27}
a=\dfrac{\pi^2}{G\Delta_1^2}<\dfrac{3}{4}
\end{equation}
As can be easily seen, the value of $y_-$, is a minimum for $y$, which should be identified with that given by Eq.(4.21), yielding
\begin{equation}\label{4-28}
y_-=-\dfrac{1}{2}-\dfrac{1}{2}\sqrt{1-\dfrac{4a}{3}}=-a=y_{min},\quad \text{or}\quad a=\dfrac{\pi^2}{G\Delta_1^2}=\dfrac{2}{3}
\end{equation}
\begin{equation}\label{4-28}
y_-=-\dfrac{2}{3},\quad y_+=-\dfrac{1}{3}
\end{equation}
Eq.(4.28) is allowed by Eq.(4.27).

Thus for a given $G>0$, we have found a fixed solution to VPT with an extremum value for $y=\ln\frac{\Delta_1^2}{\mu_s^2}=y_-=-\frac{2}{3}$ ensured by Equations
\begin{equation}\label{4-30}
\left.\dfrac{\partial E_{vac}}{\partial(\Delta^2)}\right|_{VPT}=\left.\dfrac{\partial^2 E_{vac}}{\partial(\Delta^2)^2}\right|_{VPT}=\left.\dfrac{\partial^3 E_{vac}}{\partial(\Delta^2)^3}\right|_{VPT}=0
\end{equation}
But if this is really a stable solution after VPT ? It needs a further guarantee given by the evaluation
\begin{equation}\label{4-31}
\dfrac{\partial^4 E_{vac}}{\partial(\Delta^2)^4}=-\dfrac{V}{8\pi^2}\left\{\dfrac{3}{\Delta^4}+\dfrac{G}{\pi^2\Delta^2}
\left(8\ln\dfrac{\Delta^2}{\mu_s^2}+3\right)-\dfrac{2G}{\pi^2\Delta^4}\left[\Delta^2\left(
\ln\dfrac{\Delta^2}{\mu_s^2}-1\right)+C_2\right]\right\}
\end{equation}
and the imposed condition being an inequality using Eq.(4.6) again
\begin{equation}\label{4-32}
\left.\dfrac{\partial^4 E_{vac}}{\partial(\Delta^2)^4}\right|_{VPT}=\dfrac{V}{8\pi^2\Delta_1^4}\left[1-\dfrac{G\Delta_1^2}{\pi^2}
\left(8\ln\dfrac{\Delta_1^2}{\mu_s^2}+3\right)\right]>0
\end{equation}
or, after the substitution of $\ln\frac{\Delta_1^2}{\mu_s^2}=y_1=y_-=-\frac{2}{3}$ and $a_1=\frac{\pi^2}{G\Delta_1^2}=\frac{2}{3}$,
\begin{equation}\label{4-33}
1-\dfrac{1}{a_1}(8y_-+3)=1+\dfrac{7}{3a_1}=\dfrac{9}{2}>0
\end{equation}
The inequality holds unambiguously and hence the stable vacuum after VPT with energy Eq.(4.16)
\begin{equation}\label{4-34}
E_{vac}|_{VPT}=-\dfrac{7V}{96\pi^2}\Delta_1^4=-\dfrac{7V}{96\pi^2}\mu_s^4\exp\left(-\dfrac{2\pi^2}{G\Delta_1^2}\right)=
-\dfrac{7V\mu_s^4}{96\pi^2}e^{-4/3}<0
\end{equation}
Notice that after VPT only one arbitrary constant left, either the fermion mass $\Delta=\Delta_1$ or the mass scale $\mu_s$. They are linked by Eq.(4.22) with $G$ being also fixed as
\begin{equation}\label{4-35}
G=\dfrac{3\pi^2}{2\Delta_1^2}=\dfrac{3\pi^2}{2\mu_s^2}e^{2/3}
\end{equation}
At first, we didn't know the primary interaction responsible for the coupling constant $G$ in the NJL model, Eq.(1.1). Now we begin to learn why $G$ is fixed eventually is because it could be viewed as some definite measure of how many energy will be released after the VPT accompanying with the mass ($\Delta_1$) creation of a fermion (say, electron) from its massless species in the naive vacuum. For further discussion, please see the last section and Appendix A.

\section{V. Nambu--Jona-Lasinio Transformation}
\label{sec:transformation}
\setcounter{equation}{0}
\renewcommand{\theequation}{5.\arabic{equation}}

An interesting method based on NJLVS, Eq.(3.1), was also first initiated by NJL in their first paper (see Eq.(3.15) in Ref.\cite{2}). For dealing with the Hamiltonian, Eq.(2.7), we define the creation (annihilation) operator for quasiparticle (its antiparticle) during the process of VPT as\footnotemark[1]\footnotetext[1]{The physical meaning is shown in Fig.3.}
\begin{equation}\label{5-1}
\left\{\begin{array}{l}
         \hat{\alpha}_{{\bf p}h}^\dag=U_p\hat{a}_{{\bf p}h}^\dag-hV_p\hat{b}_{-{\bf p}h} \\[3mm]
         \hat{\beta}_{-{\bf p}h}=U_p\hat{b}_{-{\bf p}h}+hV_p\hat{a}^\dag_{{\bf p}h}
       \end{array}\right.
\end{equation}
Under the space-time inversion (strong reflection), Eq.(5.1) remains invariant, since
\begin{equation}\label{5-2}
\widehat{\cal PT}\hat{\alpha}_{{\bf p}h}^\dag(\widehat{\cal PT})^{-1}=\hat{\beta}_{{\bf p}-h}=U_p\hat{b}_{{\bf p}-h}-hV_p\hat{a}^\dag_{-{\bf p}-h}
\end{equation}
coinciding with Eq.(5.1) (see Eq.(2.12)). The helicity $h$ in front of $V_p$ in Eq.(5.1) is also important to ensure its invariance under a pure space inversion since
\begin{equation}\label{5-3}
\left\{\begin{array}{l}
         \hat{U}(P)\hat{\alpha}_{{\bf p}h}^\dag\hat{U}^{-1}(P)=\hat{\alpha}_{-{\bf p}-h}^\dag=U_p\hat{a}_{-{\bf p}-h}^\dag+hV_p\hat{b}_{{\bf p}-h} \\[3mm]
         \hat{U}(P)\hat{\beta}_{-{\bf p}h}\hat{U}^{-1}(P)=-\hat{\beta}_{{\bf p}-h}=U_p(-\hat{b}_{{\bf p}-h})+hV_p\hat{a}^\dag_{-{\bf p}-h}
       \end{array}\right.
\end{equation}
coinciding with Eq.(5.1) (see Eq.(2.14)).

Based on Eqs.(2.2) and (3.5), we can prove the anticommutation relations as
\begin{equation}\label{5-4}
[\hat{\alpha}_{{\bf p}h},\hat{\alpha}_{{\bf p'}h'}^\dag]_+=[\hat{\beta}_{{\bf p}h},\hat{\beta}_{{\bf p'}h'}^\dag]_+=\delta_{\bf pp'}\delta_{hh'}
\end{equation}
By using Eq.(3.5), one can derive from Eq.(5.1) its reversed transformation as
\begin{equation}\label{5-5}
\left\{\begin{array}{l}
         \hat{a}_{{\bf p}h}^\dag=U_p\hat{\alpha}_{{\bf p}h}^\dag+hV_p\hat{\beta}_{-{\bf p}h},\,
         \hat{a}_{{\bf p}h}=U_p\hat{\alpha}_{{\bf p}h}+hV_p\hat{\beta}^\dag_{-{\bf p}h}\\[3mm]
         \hat{b}_{-{\bf p}h}=U_p\hat{\beta}_{-{\bf p}h}-hV_p\hat{\alpha}^\dag_{{\bf p}h},\,
         \hat{b}_{-{\bf p}h}^\dag=U_p\hat{\beta}_{-{\bf p}h}^\dag-hV_p\hat{\alpha}_{{\bf p}h}
       \end{array}\right.
\end{equation}
Hence Eq.(5.1) implies the relativistic canonical transformation for fermions, obviously the counterpart of nonrelativistic Bogoliubov-Valatin transformation (BVT) which works so well for BCS theory of superconductivity (Refs.\cite{24,25}, see also \cite{16} and section 8.5B in \cite{17}). We will call Eq.(5.1) the NJL transformation (NJLT), whose advantage over BVT lies in the fact that not only the momentum and angular momentum are conserved before and after both transformations, but NJLT also conserves the fermion number which is vital to the relativistic QFT (whereas BVT fails to do so).

Now we may substitute Eq.(5.5) into Eqs.(2.7)-(2.9) with (3.9)-(3.11), arranging them in terms of operators for quasiparticles in normal ordering carefully, ending up with \footnotemark[1]\footnotetext[1]{In the elementary but tedious calculation, once we encounter some specific cases where ${\bf p}\neq{\bf p'}$ and/or $h\neq h'$, (as discussed before and after Eq.(3.12)) an extra factor $(1-\delta_{\bf pp'}\delta_{hh'})$ must be added by hand. Moreover, an approximation will be made in the four operator product terms $\hat{H}_4$ (see next section).}
\begin{equation}\label{5-6}
\hat{H}=E_{vac}+\hat{H}_1+\hat{H}_2+\hat{H}_4
\end{equation}
where the vacuum energy (with no operator)
\begin{equation}\label{5-7}
E_{vac}=4\sum_{\bf p}\omega_pV_p^2-\dfrac{16G}{V}\left[\sum_{\bf p}U_pV_p\right]^2+\dfrac{8G}{V}\sum_{\bf p}U^2_pV^2_p
\end{equation}
is just Eq.(3.15) gained from NJLVS as expected. And
\begin{equation}\label{5-8}
\hspace*{-5mm}\hat{H}_1\!=\!\!\sum_{{\bf p},h}\!\left\{\omega_p(U_p^2\!-\!V_p^2)\!+\!\dfrac{16G}{V}\!\left(\sum_{\bf p'}U_{p'}V_{p'}\right)U_pV_p
\!+\!\dfrac{G}{V}[(U_p^2\!-\!V_p^2)^2\!-\!4U_p^2V_p^2]\right\}\!(\hat{\alpha}_{{\bf p}h}^\dag\hat{\alpha}_{{\bf p}h}+\hat{\beta}_{{\bf p}h}^\dag\hat{\beta}_{{\bf p}h})
\end{equation}
If $V\to\infty$ for lepton case, we ignore the last term in $\{\cdots\}$, (see discussion after Eq.(3.15)). Then after substitution of solution for VPT, Eqs.(3.18)-(3.24), into $\hat{H}_1$, we find
\begin{equation}\label{5-9}
 \hat{H}_1=\sum_{{\bf p},h}E_p(\hat{\alpha}_{{\bf p}h}^\dag\hat{\alpha}_{{\bf p}h}+\hat{\beta}_{{\bf p}h}^\dag\hat{\beta}_{{\bf p}h}),\quad (E_p=\sqrt{p^2+\Delta_1^2})
\end{equation}
as expected too. It is easy to prove that
\begin{equation}\label{5-10}
\hat{\alpha}_{{\bf p}h}|\tilde{0}\rangle=\hat{\beta}_{-{\bf p}h}|\tilde{0}\rangle=0
\end{equation}
which means that there is no any single free quasiparticle (say, $e^-$) or its antiparticle (say, $e^+$) existing in the NJLVS after VPT. Eq.(5.9) simply implies that once one of them is created by whatsoever external process, it will have energy $E_p$ with mass $\Delta_1$ (say, $m_e$). However, the $\hat{H}_2$ is more interesting:
\begin{equation}\label{5-11}
\hspace*{-3mm} \hat{H}_2\!=\!\sum_{{\bf p},h}\!\left\{2\omega_pU_pV_p\!-\!\dfrac{8G}{V}\!\left(\!\sum_{\bf p'}U_{p'}V_{p'}\!\right)(U^2_p\!-\!V^2_p)
\!+\!\dfrac{4G}{V}U_pV_p(U_p^2\!-\!V_p^2)\!\right\}\!(h\hat{\alpha}_{{\bf p}h}^\dag\hat{\beta}^\dag_{-{\bf p}h}+h\hat{\beta}_{-{\bf p}h}\hat{\alpha}_{{\bf p}h})
\end{equation}
which would imply some possibility that a pair of quasiparticle and its antiparticle could be created or annihilated spontaneously in the new vacuum. A stable NJLVS should not allow such spontaneous pair creation (annihilation) process from happening. So the "dangerous terms" containing in the $\hat{H}_2$ must be eliminated as
\begin{equation}\label{5-12}
2\omega_pU_pV_p-\dfrac{8G}{V}\left(\sum_{\bf p'}U_{p'}V_{p'}\right)(U^2_p-V^2_p)
\!+\!\dfrac{4G}{V}U_pV_p(U_p^2\!-\!V_p^2)=0
\end{equation}
The above explanation can be further justified by the stability criterion for the NJLVS as shown by Eqs.(3.16)-(3.17), now from Eq.(5.7) we have:
\begin{equation}\label{5-13}
\dfrac{\partial E_{vac}}{\partial V_p}=8\omega_pV_p-\dfrac{32G}{V}\left(\sum_{\bf p'}U_{p'}V_{p'}\right)\left(U_p-\dfrac{V^2_p}{U_p}\right)\!+\!\dfrac{16G}{V}U_pV_p\left(U_p-\dfrac{V^2_p}{U_p}\right)=0
\end{equation}
which coincides with Eq.(5.12) or (3.17) precisely as expected again ($V\to\infty$).

After NJLT, the four operator product terms are collected into $\hat{H}_4$ in Eq.(5.6) and will be discussed in the next section.

\section{VI. Scalar Boson Excited as Collective Mode of Fermion-Antifermion Pairs in NJLVS}
\label{sec:collective}
\setcounter{equation}{0}
\renewcommand{\theequation}{6.\arabic{equation}}

The Hamiltonian $\hat{H}_4\approx \hat{H}_{22}$ in Eq.(5.6) after NJLT reads
\begin{equation}\label{6-1}
\begin{array}{l}
 \hat{H}_{22}=-\dfrac{G}{V}\sum_{{\bf p,p'},h,h'}(hh')\{2(U_p^2-V_p^2)(U_{p'}^2-V_{p'}^2)\hat{\alpha}_{{\bf p}h}^\dag\hat{\beta}_{-{\bf p}h}^\dag \hat{\beta}_{-{\bf p'}h'}\hat{\alpha}_{{\bf p'}h'}\\[3mm]
 \qquad+8\delta_{\bf pp'}\delta_{hh'}(U_pV_p)^2\hat{\alpha}_{{\bf p}h}^\dag\hat{\beta}_{-{\bf p}h}^\dag \hat{\beta}_{-{\bf p}h}\hat{\alpha}_{{\bf p}h}+[(U_p^2-V_p^2)(U_{p'}^2-V_{p'}^2)\\[3mm]
 \qquad-\delta_{\bf pp'}\delta_{hh'}(U_p^2-V_p^2)^2](\hat{\alpha}_{{\bf p}h}^\dag\hat{\beta}_{-{\bf p}h}^\dag\hat{\alpha}_{{\bf p'}h'}^\dag\hat{\beta}_{-{\bf p'}h'}^\dag+\hat{\beta}_{-{\bf p}h}\hat{\alpha}_{{\bf p}h}\hat{\beta}_{-{\bf p'}h'}\hat{\alpha}_{{\bf p'}h'})\}
\end{array}
\end{equation}
where some terms like $\hat{\alpha}^\dag\hat{\beta}^\dag\hat{\alpha}^\dag\hat{\alpha}$ or $\hat{\alpha}^\dag\hat{\alpha}\hat{\beta}\hat{\alpha}$ had been erased in accordance with the condition $\hat{H}_2=0$, Eq.(5.12). The physical implication of the existence of $\hat{H}_{22}$ is as follows. The NJLVS is actually a correlated vacuum, filling with virtual pairs of massless particle and its antiparticle (say, massless $e^-e^+$), they fluctuate and transform each other via the interactions shown by Eq.(3.10) and Fig.1. Therefore, it is also possible to create a scalar boson as the collective mode composed of $f\bar{f}$ pairs (say, massive $e^-e^+$), based on the Hamiltonian $\hat{H}=\hat{H}_1+\hat{H}_{22}$ with NJLVS as its background.

To this purpose, we first define operators for "quasiboson" with zero external momentum as \footnotemark[1]\footnotetext[1]{Our early study on this topic was published in Ref.\cite{23} (see also section 8.4.3 in Ref.\cite{12}). More discussions with some corrections are presented in this paper. Notice that
\begin{equation*}
\widehat{\cal PT}\hat{B}_{{\bf p}h}^\dag(\widehat{\cal PT})^{-1}=\hat{B}_{-{\bf p}-h},\quad \widehat{\cal PT}\hat{B}_{{\bf p}h}(\widehat{\cal PT})^{-1}=\hat{B}^\dag_{-{\bf p}-h}
\end{equation*}
}
\begin{equation}\label{6-2}
\hat{B}_{{\bf p}h}^\dag=h\hat{\alpha}_{{\bf p}h}^\dag\hat{\beta}_{-{\bf p}h}^\dag,\quad \hat{B}_{{\bf p}h}=h\hat{\beta}_{-{\bf p}h}\hat{\alpha}_{{\bf p}h},\quad ({\bf p},h \quad\text{not summed})
\end{equation}
It is easy to prove that, using Eq.(5.3)
\begin{equation}\label{6-3}
\hat{U}(P)\hat{B}_{{\bf p}h}^\dag\hat{U}^{-1}(P)=\hat{B}_{-{\bf p}-h}^\dag
\end{equation}
and the commutation relation being approximately
\begin{equation}\label{6-4}
[\hat{B}_{{\bf p}h}, \hat{B}_{{\bf q}s}^\dag]=\delta_{\bf pq}\delta_{hs}-\delta_{\bf pq}\delta_{hs}(\hat{\alpha}_{{\bf p}h}^\dag\hat{\alpha}_{{\bf p}h}+\hat{\beta}_{-{\bf p}h}^\dag\hat{\beta}_{-{\bf p}h})\approx\delta_{\bf pq}\delta_{hs}
\end{equation}
(Here $s$ is the helicity, not spin projection in space). The approximation made in Eq.(6.4) bears some resemblance to that of a Cooper pair in the theory for superconductivity \cite{26} (see also \eg, section 8.5 in Ref.\cite{17}). And this approximation is complying with that in deriving Eq.(6.1). Indeed,
\begin{equation}\label{6-5}
\begin{array}{l}
 \hat{H}_{22}=-\dfrac{G}{V}\sum\limits_{{\bf p,p'},h,h'}\{[2(U_p^2-V_p^2)(U_{p'}^2-V_{p'}^2)
 +8\delta_{\bf pp'}\delta_{hh'}U_p^2V_p^2]\hat{B}_{{\bf p}h}^\dag\hat{B}_{{\bf p'}h'}\\[5mm]
\qquad +[(U_p^2-V_p^2)(U_{p'}^2-V_{p'}^2)-\delta_{\bf pp'}\delta_{hh'}(U_p^2-V_p^2)^2](\hat{B}_{{\bf p}h}^\dag\hat{B}_{{\bf p'}h'}^\dag+\hat{B}_{{\bf p}h}\hat{B}_{{\bf p'}h'})\}
\end{array}
\end{equation}
Then we construct a rest "phonon operator" \cite{23}
\begin{equation}\label{6-6}
\hat{Q}_n^{(s)\dag}=\sum_{{\bf q},s}(a_{nq}\hat{B}_{{\bf q}s}^\dag+b_{nq}\hat{B}_{{\bf q}s})
\end{equation}
for describing a scalar boson with $J^P=0^+$ since
\begin{equation}\label{6-7}
\hat{U}(P)\hat{Q}_n^{(s)\dag}\hat{U}^{-1}(P)=\hat{Q}_n^{(s)\dag}
\end{equation}
The superscript $(s)$ means "scalar" and the subscript $n$ refers to the stationary state of phonon. The coefficients $a_{nq}$ and $b_{nq}$ are waiting to be fixed.

Another "pseudo scalar" ($J^P=0^-$) phonon operator can be constructed as
\begin{eqnarray}
  &&\hat{Q}_n^{(ps)\dag}=\sum_{{\bf q}}[c_{nq}(\hat{B}_{{\bf q}1}^\dag-\hat{B}_{{\bf q}-1}^\dag)+d_{nq}(\hat{B}_{{\bf q}1}-\hat{B}_{{\bf q}-1})]=\sum_{{\bf q},s}[c_{nq}s\hat{B}_{{\bf q}s}^\dag+d_{nq}s\hat{B}_{{\bf q}s}] \\[4mm]
  &&\hat{U}(P)\hat{Q}_n^{(ps)\dag}\hat{U}^{-1}(P)=-\hat{Q}_n^{(ps)\dag}
\end{eqnarray}
To find the rest energy $\hbar\Omega_n=\Omega_n$ of such a "phonon", a method of so-called "Random Phase Approximation" (RPA) first introduced by Bohm and Pines in 1951-1953 \cite{27,28} for studying the electron gas in condensed matter physics, then used effectively in nuclear physics \cite{16,29}, seems also suitable to NJL model here. The RPA equation for Eq.(6.6) reads
\begin{equation}\label{6-10}
[\hat{H},\hat{Q}_n^{(s)\dag}]=\Omega_n^{(s)}\hat{Q}_n^{(s)\dag}
\end{equation}
For evaluating the commutator at the LHS, the following formulas are useful
\begin{equation}\label{6-11}
\left.\begin{array}{l}
[\hat{\alpha}^\dag_{{\bf p}h}\hat{\alpha}_{{\bf p}h},\hat{B}_{{\bf q}s}^\dag]=
 [\hat{\beta}_{-{\bf p}-h}^\dag\hat{\beta}_{-{\bf p}-h},\hat{B}^\dag_{{\bf q}s}]=\hat{B}_{{\bf q}s}^\dag\delta_{\bf pq}\delta_{hs}\\[3mm]
[\hat{\alpha}^\dag_{{\bf p}h}\hat{\alpha}_{{\bf p}h},\hat{B}_{{\bf q}s}]=
 [\hat{\beta}_{-{\bf p}-h}^\dag\hat{\beta}_{-{\bf p}-h},\hat{B}_{{\bf q}s}]=-\hat{B}_{{\bf q}s}\delta_{\bf pq}\delta_{hs}\\[3mm]
[\hat{B}^\dag_{{\bf p}h}\hat{B}_{{\bf p'}h'},\hat{B}_{{\bf q}s}^\dag]=\hat{B}^\dag_{{\bf p}h}\delta_{\bf p'q}\delta_{h's},
 [\hat{B}_{{\bf p}h}^\dag\hat{B}_{{\bf p'}h'},\hat{B}_{{\bf q}s}]=-\hat{B}_{{\bf p'}h'}\delta_{\bf pq}\delta_{hs}\\[3mm]
 [\hat{B}_{{\bf p}h}^\dag\hat{B}_{{\bf p'}h'}^\dag,\hat{B}_{{\bf q}s}]=-\hat{B}_{{\bf p}h}^\dag\delta_{\bf p'q}\delta_{h's}
 -\hat{B}_{{\bf p'}h'}^\dag\delta_{\bf pq}\delta_{hs} \\[3mm]
 [\hat{B}_{{\bf p}h}\hat{B}_{{\bf p'}h'},\hat{B}_{{\bf q}s}^\dag]=\hat{B}_{{\bf p}h}\delta_{\bf p'q}\delta_{h's}
 +\hat{B}_{{\bf p'}h'}\delta_{\bf pq}\delta_{hs}
\end{array}\right\}
\end{equation}
Substituting Eq.(6.6) into the RHS of Eq.(6.10) (with ${\bf q}\to {\bf p}, s\to h$) and comparing the coefficients of operators
$\hat{B}_{{\bf p}h}^\dag$ and $\hat{B}_{{\bf p}h}$ respectively, we get two coupling equations for $y_{np}=a_{np}+b_{np}$ and $x_{np}=a_{np}-b_{np}$ as
\begin{eqnarray}
  &&\hspace*{-2cm}\Omega_n^{(s)}y_{np}\!-\!2E_px_{np}\!+\!\dfrac{8G}{V}\!\left[\sum_{\bf q}(U_q^2\!-\!V_q^2)x_{nq}\right]\!(U_p^2\!-\!V_p^2)\!-\! \dfrac{2G}{V}\left[
  (U_p^2\!-\!V_p^2)\!-\!4U_p^2V_p^2\right]\!x_{np}\!=\!0\\[3mm]
  &&\hspace*{-2cm}\Omega_n^{(s)}x_{np} - 2E_py_{np}+ \dfrac{2G}{V}y_{np}=0
\end{eqnarray}
Considering the case for lepton, $V\to\infty$ but $G=\frac{3\pi^2}{2\Delta_1^2}$ finite, so $\frac{G}{V}\to 0$, we find from Eq.(6.13) that
\begin{equation}\label{6-14}
y_{np}=\dfrac{\Omega_n^{(s)}}{2E_p}x_{np}
\end{equation}
Substituting of Eq.(6.14) into Eq.(6.12) yields
\begin{equation}\label{6-15}
\dfrac{[\Omega_n^{(s)}]^2}{2E_p}x_{np}-2E_px_{np}+\dfrac{8G}{V}\left[\sum_{\bf q}(U_q^2-V_q^2)x_{nq}\right](U_p^2-V_p^2)=0
\end{equation}
Since $U_q^2-V_q^2=\frac{\omega_q}{E_q}, U_p^2-V_p^2=\frac{\omega_p}{E_p}$ (see Eq.(3.21)), we tentatively assume that $x_{np}=A\frac{\Delta_1}{\omega_p}$ with $A$ being a constant. Then Eq.(6.15) becomes
\begin{equation}\label{6-16}
[\Omega_n^{(s)}]^2=4E^2_p-4\left[\dfrac{4G}{V}\left(\sum_{\bf q}\dfrac{1}{E_q}\right)\right]\omega_p^2=4\Delta^2_1
\end{equation}
where Eq.(3.23) has been used. Hence Eq.(6.10) is solved as
\begin{equation}\label{6-17}
\Omega_n^{(s)}=2\Delta_1
\end{equation}
implying that the $0^+$ boson, being a collective mode of numerous (indefinite) pairs $f\bar{f}$, just has the mass of $2\Delta_1=m_f+m_{\bar{f}}=2m_f$. The existence of bosons as collective modes of $f\bar{f}$ pairs was first predicted by Nambu in the BCS theory\cite{30} and Eq.(6.17) was first calculated in NJL model\cite{2} by the use of the Bethe-Salpeter equation. Of course, the result will be dramatically different if we consider three leptons coupling together in a future study.

Now we turn to the solution of Eq.(6.8) and
\begin{equation}\label{6-18}
[\hat{H}, \hat{Q}_n^{(ps)\dag}]=\Omega_n^{(ps)}\hat{Q}_n^{(ps)\dag}
\end{equation}
for $0^-$ boson. Define $X_{np}=c_{np}-d_{np},Y_{np}=c_{np}+d_{np}$, instead of Eqs.(6.12)-(6.15), we find
\begin{eqnarray}
  \Omega_n^{(ps)}Y_{np} &=& 2E_pX_{np}+\dfrac{2G}{V}[(U_p^2-V_P^2)^2-4U_p^2V_p^2]X_{np} \\[4mm]
  \Omega_n^{(ps)}X_{np} &=& 2E_pY_{np}-\dfrac{2G}{V}Y_{np}
\end{eqnarray}
Notice that a term like that in Eq.(6.12) disappears due to $\sum_{\bf q}(U_q^2-V_q^2)(c_{nq}-c_{nq})=0$ \etc ~and the last term in Eq.(6.19) or (6.20) vanishes because $V\to\infty$ (for leptons), ending up with
\begin{equation}
\begin{array}{l}
  \Omega_n^{(ps)}Y_{np} = 2E_pX_{np} \\[4mm]
  \Omega_n^{(ps)}X_{np} = 2E_pY_{np}
\end{array}
\end{equation}
To solve Eq.(6.21), we consider the following formulas
\begin{equation}
\sum_{\bf p}\dfrac{1}{E_p}=\dfrac{V}{4G}\qquad (\text{see Eq.(3.23)})
\end{equation}
\begin{eqnarray}
  \sum_{\bf p}\dfrac{1}{E^2_p} &=& \dfrac{V}{2\pi^2}[-\dfrac{\pi}{2}(\Delta_1-F)]\quad (\text{with $F$ being an arbitrary constant}) \\[4mm]
  \sum_{\bf p}\dfrac{1}{E^3_p} &=& \dfrac{V}{2\pi^2}(-\dfrac{1}{2}\ln\dfrac{\Delta_1^2}{\mu_s^2})=-\dfrac{V}{4\pi^2}y_1=\dfrac{V}{6\pi^2} \\[4mm]
  \sum_{\bf p}\dfrac{1}{E^4_p} &=& \dfrac{V}{8\pi\Delta_1},\quad  \sum_{\bf p}\dfrac{1}{E^5_p}=\dfrac{V}{6\pi^2\Delta^2_1},\ldots
\end{eqnarray}
Obviously, $\sum_{\bf p}\dfrac{1}{E^{n+3}_p}$ is a fixed quantity (apart from the volume $V$) as long as the integer $n\geq0$. So we may assume, \eg, $n=2$ that
\begin{equation}\label{6-25}
Y_{2p}=\dfrac{\Delta_1^5}{E_p^5},\quad X_{2p}=\xi_2\dfrac{\Delta_1^5}{E_p^5}
\end{equation}
For convenience here, we denote the subscript $"n"$ as the "order" of approximation in calculation. However, the result in this case will be independent of the exact value of $n$ as long as $n\geq0$.

Substitution of Eq.(6.26) into Eq.(6.21) and the summation over ${\bf p}$ lead to
\begin{eqnarray}
  \Omega_2^{(ps)}\sum_{\bf p}\dfrac{1}{E^5_p} &-& 2\xi_2\sum_{\bf p}\dfrac{1}{E^4_p}=0 \\[4mm]
  \xi_2\Omega_2^{(ps)}\sum_{\bf p}\dfrac{1}{E^5_p} &-& 2\sum_{\bf p}\dfrac{1}{E^4_p}=0
\end{eqnarray}
Subtraction of $\xi_2\times$Eq.(6.28) from Eq.(6.27) yields
\begin{equation}\label{6-28}
(1-\xi_2^2)\Omega_2^{(ps)}\sum_{\bf p}\dfrac{1}{E^5_p}=(1-\xi_2^2)\Omega_2^{(ps)}\dfrac{V}{6\pi^2\Delta^2_1}=0
\end{equation}
which implies two solutions (with generalization $2\to n$)
\begin{eqnarray}
  \text{Either}\quad \Omega_n^{(ps)} &=& 0\quad \text{with}\quad \xi_n\neq1,\quad (n\geq0) \\
  \text{or}\quad \Omega_n^{(ps)} &\neq & 0\quad \text{with}\quad \xi_n=1,\quad (n\geq0)
\end{eqnarray}
Eq.(6.30) just means the Goldstone boson predicted first by Nambu \cite{30} in the context of the BCS superconductivity and by Goldstone for the QFT model in which spinless particles of zero mass always happen when a continuous symmetry group leaves the Lagrangian but not the vacuum invariant \cite{31}. This "Goldstone's conjecture" was systematically generalized and proved by Goldstone, Salam and Weinberg \cite{32} into a theorem that "if there is continuous symmetry transformation under which the Lagrangian is invariant, then either the vacuum state is also invariant under the transformation, or there must exist spinless particles of zero mass." Here Eq.(6.30) reconfirms that a Goldstone boson with $J^P=0^-$ and mass $\Omega^{(ps)}=0$ must be a unique collective mode after VPT. It is composed of unfixed (actually infinite) numbers of massive lepton-antilepton pairs as shown by Eqs.(6.1)-(6.9) and (6.26) because $\xi_n=\frac{c_{np}-d_{np}}{c_{np}+d_{np}}\neq1$ means $\frac{d_{np}}{c_{np}}\neq0$ ---the creation and annihilation of pairs are happening unceasingly in a Goldstone boson such that its zero mass can be ensured.

In nonrelativistic condensed matter physics, a Goldstone boson can be observed, \eg, as the phonon excitation in the superfluid ${}^4He$ as a common consequence of spontaneous breaking of continuous Galilean symmetry and that of condensed number of zero momentum ${}^4He$ atoms --- the dispersion relation $\omega(k)$ has no gap at $k\to0$. Similarly, the spin waves (magnons) in magnet can be viewed as Goldstone bosons due to the original continuous rotational symmetry of magnetization direction being spontaneously broken (see, \eg, \cite{33}).

However, in relativistic QFT, once continuous phase symmetries are coupled with gauge fields, say, in the $SU(2)\times U(1)$ electroweak model, the three would-be Goldstone bosons are "eaten up" by three gauge bosons, $W^\pm$ and $Z$, becoming the latters' longitudinal polarization degrees of freedom and thus $W^\pm$ and $Z$ acquire their masses. This Higgs mechanism was first discovered by Higgs\cite{34}, Englert and Brout\cite{35}, Guralnik, Hagen and Kibble \cite{36} simultaneously in 1964. Please see also Ref.\cite{12}.

Then what does Eq.(6.31) mean? To our understanding, it corresponds to, say, the familiar ground state of positronium, so-called "para-positronium", which has $J^P=0^-$ too (see Eq.(B.4)). Another "ortho-positronium" has $J^P=1^-$. Their average energy can be well described by
\begin{equation}\label{6.35}
E_{aver.}(n)=2m_e-\dfrac{m_e\alpha^2}{4n^2}
\end{equation}
where $n$ is the principal quantum number. But the energy difference between them has been measured experimentally \cite{37} [with notation $E(n{}^{2S+1}L_J)$]:
\begin{equation}\label{3-36}
\Delta E_{1S}=E_{ortho}-E_{para}=E(1{}^3S_1)-E(1{}^3S_0)=203389.10(0.74)\;MHz
\end{equation}
Theoretically, this hyperfine structure (HFS) of positronium $1S$ states was first derived by Karplus and Klein in 1952 \cite{38} to be
\begin{equation}\label{6.37}
\Delta E_{1S}=\dfrac{7}{12}m_e\alpha^4-\dfrac{m_e\alpha^5}{\pi}(\dfrac{8}{9}+\dfrac{1}{2}\ln 2)
\end{equation}
and later by various authors to high accuracy \cite{39,40,41}. To our understanding, even though the leading coefficient $\frac{7}{12}$ is composed of two contributions from both the scattering and annihilation channels
\begin{equation}\label{6.38}
\dfrac{7}{12}=(\dfrac{1}{3})_{scatt}+(\dfrac{1}{4})_{annih}
\end{equation}
only one pair of $(e^+e^-)$ configuration is involved. And this is in conformity with Eq.(6.31) that once $\Omega(\text{para-positronium})\neq0$, it must have $\xi=1$, so not a collective mode. We will discuss boson with $J^P=1^-$ like ortho-positronium in the next section. The $C$ (charge) parity will be discussed in the Appendix B.

\section{VII. Vector Boson Excited as Collective Mode after VPT}
\label{sec:discussion}
\setcounter{equation}{0}
\renewcommand{\theequation}{7.\arabic{equation}}

For discussing vector bosons with $J^P=1^+$ and $1^-$, instead of the "quasiboson" defined at Eq.(6.2), we define
\begin{equation}\label{7-1}
\hat{\cal B}_{{\bf p}h}^\dag=h\hat{\alpha}_{{\bf p}h}^\dag\hat{\beta}_{-{\bf p}-h}^\dag,\quad \hat{\cal B}_{{\bf p}h}=h\hat{\beta}_{-{\bf p}-h}\hat{\alpha}_{{\bf p}h}\quad ({\bf p},h \, \text{not summed})
\end{equation}
Note that $\hat{\beta}$ has helicity $"-h"$, so $\hat{\cal B}$ has zero external momentum but its "spin" $J=1$. Similar to Eqs.(2.11) and (6.3), now we have
\begin{eqnarray}
  \widehat{\cal PT}\hat{\cal B}_{{\bf p}h}^\dag(\widehat{\cal PT})^{-1} &=& -\hat{\cal B}_{-{\bf p}h} \\
  \hat{U}(P)\hat{\cal B}_{{\bf p}h}^\dag\hat{U}^{-1}(P) &=& \hat{\cal B}_{-{\bf p}-h}^\dag
\end{eqnarray}
Also similar to Eqs.(6.11) and (6.4), we have
\begin{equation}\label{7-4}
 \begin{array}{l}
  [\hat{\alpha}_{{\bf p}h}^\dag\hat{\alpha}_{{\bf p}h}, \hat{\cal B}_{{\bf q}s}^\dag]=[\hat{\beta}_{-{\bf p}-h}^\dag\hat{\beta}_{-{\bf p}-h},
\hat{\cal B}_{{\bf q}s}^\dag]=\delta_{\bf pq}\delta_{hs}\hat{\cal B}_{{\bf q}s}^\dag\\[3mm]
[\hat{\alpha}_{{\bf p}h}^\dag\hat{\alpha}_{{\bf p}h}, \hat{\cal B}_{{\bf q}s}]=[\hat{\beta}_{-{\bf p}-h}^\dag\hat{\beta}_{-{\bf p}-h}, \hat{\cal B}_{{\bf q}s}]=-\delta_{\bf pq}\delta_{hs}\hat{\cal B}_{{\bf q}s}
 \end{array}
\end{equation}
\begin{equation}\label{7-5}
[\hat{\cal B}_{{\bf p}h},\hat{\cal B}_{{\bf q}s}^\dag]=\delta_{\bf pq}\delta_{hs}
\end{equation}
Look carefully at Eq.(6.1), we find that only one kind of terms in $\hat{H}_{22}$ can be recast into that expressed by $\hat{\cal B}_{{\bf p}h}$, they are
\begin{equation}\label{7-6}
\begin{array}{l}
\hat{H}'_{22}=-\dfrac{G}{V}\sum\limits_{{\bf p,p'},h,h'}\{hh'(U_p^2-V_p^2)(U_{p'}^2-V_{p'}^2)(\hat{\alpha}_{{\bf p}h}^\dag\hat{\beta}_{-{\bf p}h}^\dag\hat{\alpha}_{{\bf p'}h'}^\dag\hat{\beta}_{-{\bf p'}h'}^\dag+\hat{\beta}_{-{\bf p}h}\hat{\alpha}_{{\bf p}h}\hat{\beta}_{-{\bf p'}h'}\hat{\alpha}_{{\bf p'}h'})\}_{{\bf p}'={\bf p}, h'=-h}\\[4mm]
=\dfrac{G}{V}\sum\limits_{{\bf p},h}(U_p^2-V_p^2)^2(\hat{\cal B}_{{\bf p}h}^\dag\hat{\cal B}_{{\bf p}-h}^\dag+\hat{\cal B}_{{\bf p}-h}\hat{\cal B}_{{\bf p}h})
\end{array}
\end{equation}
where ${\bf p}'={\bf p}$ but $h'=-h$ are set to ensure the conservation of both momentum and angular momentum in the vacuum.

Similar to but different from Eqs.(6.6) and (6.8), we define a pseudo-vector boson operator with $J^P=1^+$ as
\begin{eqnarray}
  \hat{Q}_n^{(pv)\dag}=\sum_{|{\bf q}|=q}[a'_{nq}\hat{\cal B}_{{\bf q}s}^\dag+b'_{nq}\hat{\cal B}_{{\bf q}s}]\\
\hat{U}(P)\hat{Q}_n^{(pv)\dag}\hat{U}^{-1}(P)=\hat{Q}_n^{(pv)\dag}
\end{eqnarray}
and the operator for vector boson with $J^P=1^-$ as:
\begin{eqnarray}
  \hat{Q}_n^{(v)\dag}=\sum_{|{\bf q}|=q}[c'_{nq}(\hat{\cal B}_{{\bf q}1}^\dag-\hat{\cal B}_{{\bf q}-1}^\dag)+d'_{nq}(\hat{\cal B}_{{\bf q}1}-\hat{\cal B}_{{\bf q}-1})]\\
\hat{U}(P)\hat{Q}_n^{(v)\dag}\hat{U}^{-1}(P)=-\hat{Q}_n^{(v)\dag}
\end{eqnarray}
Notice that only $|{\bf q}|=q$ (not its direction) is summed in both Eqs.(7.7) and (7.9). To solve the RPA equation for boson defined by Eq.(7.7)
\begin{equation}\label{7-11}
[\hat{H}',\hat{Q}_n^{(pv)\dag}]=\Omega_n^{(pv)}\hat{Q}_n^{(pv)\dag}
\end{equation}
where
\begin{equation}\label{7-12}
\begin{array}{l}
 \hat{H}'=\hat{H}_1+\hat{H}'_{22} \\
\hat{H}_1=\sum_{{\bf p},h}E_p(\hat{\alpha}_{{\bf p}h}^\dag\hat{\alpha}_{{\bf p}h}+\hat{\beta}_{-{\bf p}-h}^\dag\hat{\beta}_{-{\bf p}-h}),\quad (E_p=\sqrt{p^2+\Delta^2_1})
\end{array}
\end{equation}
we define $y'_{np}=a'_{np}+b'_{np},x'_{np}=a'_{np}-b'_{np}$ and find
\begin{eqnarray}
  \Omega_n^{(pv)}y'_{np} &=&2E_px'_{np}+\dfrac{2G}{V}(U_p^2-V_p^2)^2x'_{np}\\[4mm]
\Omega_n^{(pv)}x'_{np} &=&2E_py'_{np}-\dfrac{2G}{V}(U_p^2-V_p^2)^2y'_{np}
\end{eqnarray}
For lepton case, $G/V\to0\,(V\to\infty)$ we have
\begin{equation}\label{7-15}
\left\{
  \begin{array}{ll}
   \Omega_n^{(pv)}y'_{np}-2E_px'_{np}=0\\
   \Omega_n^{(pv)}x'_{np}-2E_py'_{np}=0
  \end{array}
\right.
\end{equation}
Just like Eqs.(6.19)-(6.30) for boson with $J^P=0^-$, we assume
\begin{equation}\label{7-16}
y'_{np}=\dfrac{\Delta_1^{n+3}}{E_p^{n+3}},\quad x'_{np}=\zeta_n^{(pv)}\dfrac{\Delta_1^{n+3}}{E_p^{n+3}},\quad (n\geq0)
\end{equation}
and reach \footnotemark[1]\footnotetext[1]{Different from Eqs.(6.23)-(6.25), here $\sum_p\frac{1}{E_p^2}=\frac{L}{2\pi}(\frac{\pi}{\Delta_1}),\sum_p\frac{1}{E_p^3}=\frac{L}{2\pi}(\frac{2}{\Delta^2_1})$, \etc, $L$ is the length in one-dimensional space. But we don't need their concrete forms in this section for leptons ($L\to\infty$). They will be used only for quarks where $L$ is finite.}
\begin{equation}\label{7-17}
\Omega_n^{(pv)}[1-(\zeta_n^{(pv)})^2]\sum_p\dfrac{1}{E_p^{n+3}}=0
\end{equation}
which implies that
\begin{eqnarray}
  \text{Either}\quad \Omega_n^{(pv)} &=& 0\quad \text{with}\quad \zeta_n^{(pv)}\neq1,\quad (n\geq0) \\
  \text{or}\quad \Omega_n^{(pv)} &\neq & 0\quad \text{with}\quad \zeta_n^{(pv)}=1,\quad (n\geq0)
\end{eqnarray}
While Eq.(7.18) refers to a Goldstone boson with $J^P=1^+$ like that with $J^P=0^-$ (see Ref.\cite{33}). Eq.(7.19) may be compared with the para-positronium (with $J^P=0^-$) discussed at the end of last section.

A vector boson operator with $J^P=1^-$ is defined at Eq.(7.9) and its motion equation
\begin{equation}\label{7-20}
[\hat{H}',\hat{Q}_n^{(v)\dag}]=\Omega_n^{(v)}\hat{Q}_n^{(v)\dag}
\end{equation}
or
\begin{equation}
\left\{
\begin{array}{l}
\Omega_n^{(v)}Y'_{np} =2E_pX'_{np}-\dfrac{2G}{V}(U_p^2-V_p^2)^2X'_{np}\\[4mm]                                                         \Omega_n^{(v)}X'_{np} =2E_pY'_{np}+\dfrac{2G}{V}(U_p^2-V_p^2)^2Y'_{np}
\end{array}
\right.
\end{equation}
where
\begin{equation}\label{7-22}
Y'_{np}=c'_{np}+d'_{np},\quad X'_{np}=c'_{np}-d'_{np}
\end{equation}
Consider lepton case, $V\to\infty$ and assume
\begin{equation}\label{7-23}
Y'_{np}=\dfrac{\Delta_1^{n+3}}{E_p^{n+3}},\quad X'_{np}=\zeta_n^{(v)}\dfrac{\Delta_1^{n+3}}{E_p^{n+3}},\quad (n\geq0)
\end{equation}
The solution reads (just like that for $J^P=1^+$ case, Eqs.(7.16)-(7.19))
\begin{eqnarray}
  \text{Either}\quad \Omega_n^{(v)} &=& 0\quad \text{with}\quad \zeta_n^{(v)}\neq1,\quad (n\geq0) \\
  \text{or}\quad \Omega_n^{(v)} &\neq & 0\quad \text{with}\quad \zeta_n^{(v)}=1,\quad (n\geq0)
\end{eqnarray}
Again, Eq.(7.24) describe another Goldstone boson with $J^P=1^-$ whereas Eq.(7.25) refers to, say, an ortho-positronium's $S$ state $(n{}^{2S+1}L_J)$ with $J=1,L=0,S=1$ and principal quantum number $n=1$, so $J^P=1^-$ too (see Eq.(B.4)).

\section{VIII. Summary and Discussions}
\label{sec:discussion}
\setcounter{equation}{0}
\renewcommand{\theequation}{8.\arabic{equation}}

1. As first pointed out by NJL\cite{2} that Eq.(1.1) is invariant under the continuous chiral transformation ($\gamma_5=\begin{scriptsize}\left(                                                                                                             \begin{array}{cc} 0 & I \\  I & 0 \\   \end{array} \right)\end{scriptsize}, \gamma_0=\begin{scriptsize}\left(                                                                                                             \begin{array}{cc} I & 0 \\  0 & -I \\   \end{array} \right)\end{scriptsize}$ in B-D metric)
\begin{equation}\label{8-1}
\psi\to e^{i\alpha\gamma_5}\psi
\end{equation}
because
\begin{equation}\label{8-2}
\left\{
  \begin{array}{ll}
   \bar{\psi}\psi\to \bar{\psi}\psi\cos2\alpha+i\bar{\psi}\gamma_5\psi\sin2\alpha\\[3mm]
    i\bar{\psi}\gamma_5\psi\to -\bar{\psi}\psi\sin2\alpha+i\bar{\psi}\gamma_5\psi\cos2\alpha
  \end{array}
\right.
\end{equation}
and so a mass term like "$-m\bar{\psi}\psi$" is forbiddon in the Lagrangian, Eq.(1.1), whereas the interaction term $G[(\bar{\psi}\psi)^2-(\bar{\psi}\gamma_5\psi)^2]$ remains invariant. The consequence of NJL model turns out to be a mass creation process of fermion which breaks the chiral symmetry.

What we wish to prove in this paper is: The chiral symmetry at the level of QFT means a product of two symmetries --- one is the discrete symmetry of "strong reflection", \ie, the (newly defined) space-time inversion (${\bf x}\to-{\bf x},t\to-t$) invariance, or the particle-antiparticle symmetry ${\cal PT}={\cal C}$ (as discussed in Ref.\cite{20}), the another is the continuous phase $\alpha$ transformation reflected in Eqs.(8.1)-(8.2).

At the level of QFT, $\psi\to\hat{\psi}$, it evolves into the field operator for fermion (see \cite{20})
\begin{eqnarray}
 \hat{\psi}({\bf x},t)\to(\widehat{\cal PT})\hat{\psi}({\bf x},t)(\widehat{\cal PT})^{-1}=\gamma_5\hat{\psi}(-{\bf x},-t)=\hat{\psi}({\bf x},t)\\
\hat{\psi}(-{\bf x},-t)=\gamma_5 \hat{\psi}({\bf x},t)
\end{eqnarray}
So the appearance of $\gamma_5$ matrix in Eqs.(8.1)-(8.2), to our understanding, implies essentially a transformation between a fermion and its antifermion, say, $e^-$ and $e^+$ (see Eqs.(5.12)-(5.15) in Ref.\cite{20}). The wonderful feature of NJL model lies in the fact that ${\cal PT}={\cal C}$ symmetry remains intact with mass creation before and after a VPT.

Actually, as the counterpart of Eq.(2.1), we have $\hat{\psi}\to\hat{\Psi}$ after the VPT and
\begin{equation}\label{8-5}
\hat{\Psi}({\bf x},t)=\dfrac{1}{\sqrt V}\sum_{{\bf p},h}\sqrt{\dfrac{\Delta_1}{E_p}}[\hat{\alpha}_{{\bf p}h}u_h({\bf p})e^{i({\bf p}\cdot{\bf x}-E_pt)}+\hat{\beta}^\dag_{{\bf p}h}v_h({\bf p})e^{-i({\bf p}\cdot{\bf x}-E_pt)}]
\end{equation}
for free field ($E_p=\sqrt{p^2+\Delta_1^2}$) with operators $\hat{\alpha},\hat{\beta}$ being defined at Eqs.(5.1)-(5.5). Following Chang-Chang-Chou in Ref.\cite{22}, we define the vacuum expectation value (VEV) of composite field operators in Eq.(8.2) as (with $''\,\ {\widehat{}}\,\ ''$ erased for clarity)
\begin{equation}\label{8-6}
\langle\bar{\psi}\psi\rangle\equiv\dfrac{1}{V}\int d{\bf x}\langle\text{vac}|\bar{\psi}\psi|\text{vac}\rangle, \quad \langle\bar{\psi}\gamma_5\psi\rangle\equiv\dfrac{1}{V}\int d{\bf x}\langle\text{vac}|\bar{\psi}\gamma_5\psi|\text{vac}\rangle
\end{equation}
Obviously, before the VPT, $|\text{vac}\rangle=|0\rangle$, the naive vacuum, both VEVs in Eq.(8.6) vanish:
\begin{equation}\label{8-7}
\langle0|\bar{\psi}\psi|0\rangle=\langle0|\bar{\psi}\gamma_5\psi|0\rangle=0
\end{equation}
(As a rule in QFT, the normal ordering has been taken for quadratic forms like $\bar{\psi}\psi$ \etc, see \cite{20}). However, after VPT, $|\text{vac}\rangle=|\tilde0\rangle$, the NJLVS, we have
\begin{equation}\label{8-8}
\langle\tilde0|\bar{\psi}\psi|\tilde0\rangle= \langle\tilde0|\bar{\Psi}\Psi|\tilde0\rangle=-\dfrac{\Delta_1^3}{3\pi^2}
\end{equation}
but
\begin{equation}\label{8-9}
\langle\tilde0|\bar{\psi}\gamma_5\psi|\tilde0\rangle= \langle\tilde0|\bar{\Psi}\gamma_5\Psi|\tilde0\rangle=0
\end{equation}
(The detail of calculations is given at the Appendix C). Hence the difference between Eq.(8.8) and (8.7) can be regarded as a "signature" of VPT being happened or not.

Returning back to Eqs.(8.1)-(8.2), we see the continuous chiral symmetry transformation is blocked at the QFT level by the non-zero VEV of $\bar{\psi}\psi$ as shown by Eq.(8.8) which is exactly the "signature" of the occurrence of VPT. However, as we will discuss later, the existence of this nonzero signature doesn't always mean the vacuum (after VPT) becoming a nonunique one.

To sum up, we may say that the mass creation mechanism provided by the NJL model strongly supports the validity of "strong reflection" or ${\cal PT}={\cal C}$ invariance in Ref.\cite{20}, or vice versa. We always stay at one inertial frame and check a theory being relativistic or not by its invariance under a space-time inversion (${\bf x}\to -{\bf x}, t\to -t$) or a mass inversion ($m\to -m$) being held or not. So this discrete symmetry is easier to use than the continuous symmetry of Lorentz transformation (among infinite inertial frames in relative motions but along one direction). This is just because ${\cal PT}={\cal C}$ is deeply rooted at the level of QFT, reflecting the symmetry between a particle and its antiparticle.

2. A new trick in this paper is the proposal of an effective Hamiltonian $\hat{H}$ for VPT shown by Eqs.(2.7)-(2.9), which considerably simplifies the calculation in NJL model. Essentially, based on $\hat{H}$, it is straightforward to evaluate the VEV of energy on the ansatz of NJLVS, Eq.(3.1). Moreover, to "diagonalize" the $\hat{H}$ by the NJLT, Eq.(5.1), becomes a systematic work, though a little tedious, it is quite fruitful. We wish to emphasize two points: (a) This $\hat{H}$ for VPT is based on the general principle of QFT, the strong reflection invariance, \ie, the ${\cal PT}={\cal C}$ symmetry \cite{20}, which should be strictly respected before and after VPT. (b) Eq.(2.9) can be modified into the quark case where the volume becomes limited or generalized into the case containing more species (like 3 leptons). The relevant researches are currently in progress.

3. However, the reason why all calculations in this paper can be simplified and getting rid of ambiguity is because we adopt a simple RRM as discussed in section IV. We first learned the concept of "divergence" from mathematical professors (when we were freshmen at univeraities) as follows: Consider a number series like $a_1,a_2,\ldots,a_n,\ldots$ and if this series is divergent, \ie, it converges into "infinity" ("$\infty$"). Then for any given large number $A$, we can always find a large integer $N$ such that if $n>N$, we have $a_n>A$. In retrospect, the above statement implies three points: (a) The divergence is meaningful only if it is relevant to numbers without any physical dimensions. (b) The latter tend to large numbers. (c) But these numbers involved are uncertain.

However, since then, when we were bothered by the "divergence" a lot in studying the QFT, we used to pay too much attention to the point (b) while overlooking points (a) and (c) to some extent. Only after adopting the RRM first proposed by one of us (Yang) in 1994, did we gradually realize that it is the point (c) that is of the essential importance to a "divergence". For example, the integral $I$ in the "gap equation", Eq.(3.24), is quadratic divergent because the upper bound of integration tends to infinity (by contrast, a similar integral in the BCS theory for superconductivity is convergent because its upper bound is cutoff naturally by the Debye frequency in the metal crystal. See \eg, Eq.(8.5.35) in Ref.\cite{17}). But after the treatment from Eqs.(4.1) through (4.5), Eq.(3.24) becomes
\begin{equation}\label{8-10}
4I=[\Delta^2(\ln\dfrac{\Delta^2}{\mu_s^2}-1)+C_2]=\dfrac{2\pi^2}{G}
\end{equation}
with two arbitrary constants $\mu_s$ and $C_2$. In our opinion, such kind of "regularization" is precisely grasping the essential meaning of a "divergence" --- the "uncertainty" in physics. Then the "renormalization" procedure for fixing $\mu_s$ and $C_2$ is also a physical condition that the new vacuum after VPT must be a stable one, see section IV. Accordingly, relevant constants are fixed as: \begin{equation}\label{8-11}
\Delta\to\Delta_1, a_1=\dfrac{\pi^2}{G\Delta_1^2}=\dfrac{2}{3}, \ln\dfrac{\Delta_1^2}{\mu_s^2}=-\dfrac{2}{3}, \dfrac{C_2}{\Delta_1^2}=3, \dfrac{I}{\Delta_1^2}=\dfrac{1}{3}
\end{equation}
Indeed, all dimensionless numbers involved are not large ones. For further discussion, please see Appendix A.

4. Based on classical electrodynamics (CED) and the theory of special relativity, the mass creation mechanism for an electron was first discussed as follows: Assuming that an electron's electric charge ($-e,e>0$) is spreading over a small sphere with radius $r_e$, one could easily estimate the electron's rest mass $m\sim \frac{e^2}{c^2r_e}$ (up to a constant depending on the unit and the charge density distribution function), which tends to infinity if $r_e\to0$, really a bad situation of divergence. Since then, the origin of mass remains as a puzzle in physics. After the invention of quantum electrodynamics (QED), physicists could calculate the "mass modification" $\delta m$ on an electron via the one loop approximation of "self-energy diagram", finding the divergence difficulty being eased into logarithmical one: $\delta m\sim -\frac{\alpha m}{4\pi}\ln\frac{\Lambda^2}{m^2}$, where $\alpha=\frac{e^2}{4\pi\hbar c}=\frac{1}{137}$ and $\Lambda$ is the cutoff in momentum integration. However, in our opinion, a better treatment on $\delta m$ is given at Ref.\cite{42}, using our RRM and arriving at
\begin{equation}\label{8-12}
\delta m= \dfrac{\alpha m}{4\pi}(5-3\ln\dfrac{m^2}{\mu_2^2})
\end{equation}
where $\mu_2$ is an arbitrary mass-scale (like $\mu_s$ in this paper). The renormalization amounts to reconfirm the $m$ in Eq.(8.12) being the mass observed experimentally: $m=m_{obs}=m_e$ when the free electron is moving on the "mass shell ($p^2=m^2$)". So the physical condition $\delta m=0$ leads to $\ln\frac{m^2}{\mu_2^2}=\frac{5}{3}$, which in turn fixes the renormalization factor for the wave function
\begin{equation}\label{8-13}
Z_2=1-\dfrac{\alpha}{3\pi}
\end{equation}
(at one-loop approximation). Now everything is finite and fixed in all finite-loop calculations of QED as shown in Refs.\cite{42,43}. Because the coupling constant $\alpha$ in QED is a dimensionless number, the perturbative calculation of "radiative corrections" cannot modify the rest mass $m$ of an electron even a bit essentially, but renders the $\alpha$ a "running coupling constant" as a function of momentum transfer in the colliding of electrons. Hence the "renormalization" of perturbative QED (p-QED) is actually a procedure of "reconfirmation" of constants $m$ and $\alpha$ in every order of finite loop approximation, \footnotemark[1]\footnotetext[1]{As a metaphor, one has to reconfirm his plane ticket before his departure from the airport. he must use the same name throughout his entire journey\cite{42,43}.} but based on a reasonable RRM in p-QED, various predictions like the anomalous magnetic moment of electron and the Lamb shift \etc, can be calculated.

So the "mass-origin puzzle" persisted until the discovery of NJL model, in which it is shown that an electron's mass is by no means created from its "self-energy". Rather, it is determined by its environment. An electron acquires its mass $m(=\Delta_1)$ only after an abrupt environment change --- after VPT, two (not one) "mass scales", $\Delta_1$ and the "signature" $<\bar{\psi}\psi>$, emerge simultaneously from zero to nonzero ones with their ratio being finite and fixed
\begin{equation}\label{8-14}
 \dfrac{\Delta_1}{<\bar{\psi}\psi>}=-\dfrac{3\pi^2}{\Delta_1^2}=-2G
\end{equation}
To our understanding, such a mass creation mechanism is only possible in a nonperturbative QFT (non p-QFT) treatment (with loop number $L\to\infty$) like NJLVS or NJLT in NJL model here. A similar model was also proposed by Gross and Neveu in 1974\cite{43}. \footnotemark[2]\footnotetext[2]{As a metaphor, one cannot lift himself from the floor even a bit by merely pulling his hair upward. By contrast, he can jump high from the floor by transferring a finite momentum impulse to the Earth --- that's the way a rocket is launched. Please see also Ref.\cite{12}.} The difference between p-QFT and non p-QFT shows up as a requirement of principle of relativity in physics and in epistemology in general \cite{45}.

To our experience, for either p-QFT or non p-QFT, our RRM can be used in a simple and flexible manner, ending up with "no explicit divergence, no counter-term,no bare parameter,and no arbitrarily running mass scale left".

5. By using the method of RPA, we try to calculate bosons as collective modes of $\bar{f}f$ pairs emerged after the VPT. For lepton case, the mass of boson with $J^P=0^+$ is found at Eq.(6.17). However, for bosons with $J^P=0^-, 1^+$ and $1^-$, we find their masses being zero as long as they are collective modes after VPT, \ie, they are Goldstone bosons. To our understanding, this is just the one case predicted by the Goldstone theorem as quoted after Eq.(6.30) from the abstract of Ref.\cite{32}. Interesting thing is: Authors of Ref.\cite{32} predicted, alternatively, that the vacuum state (after VPT) may still survive the continuous transformation, \ie, remains as an invariant ground state for QFT and then all Goldstone bosons will be gone. Why? A careful examination on their proofs \cite{32} reveals that a necessary condition for the appearance of Goldstone boson is the volume $V$ approaching infinity in the case, say, for leptons (see \cite{33}). Once if $V$ is finite, so is the vacuum energy. Thus different vacuum states labelled by the continuous phase angle $\alpha$ in Eqs.(8.1)-(8.2) will be mixed up again into one unique vacuum state after VPT, just like what happens for the ground state in QM. (see the example of one particle moving in two coupled potential wells in one space dimension discussed, \eg, in Eqs.(3.17)-(3.18) of Ref.\cite{17}). We guess the later case will be realized for the quark confined in a limited volume and so the $SU(3)_c$ symmetry in QCD will be preserved after the VPT, where there will be no zero-mass Goldstone boson and hence no Higgs mechanism as well --- the gluon remains massless.

6. Another conjecture could be as follows: While the fermion is qualified as an elementary particle in the sense of it acquiring mass via a VPT described by the NJL model, a boson seems unlikely an elementary one, as doubted by many physicists (see relevant discussions in Chapter 22 of Ref.\cite{46}). We share the same feeling. And in particular, the Higgs boson with $J^P=0^+$ could be a collective mode of quark-antiquark pairs with $t\bar{t}$ as its main configuration, as pointed out by T. D. Lee in 2007 \cite{47}.

\section*{Appendix A: Dimensionless Evaluation for Renormalization in NJL Model for Lepton Case}

Let us recalculate Eqs.(4.12)-(4.35) in dimensionless form so that they can be easily handled by the computer and generalized to later study, \eg, for quark case. Introduce the dimensionless square of running order parameter $x=\Delta^2/\Delta_1^2$ and the density of vacuum energy
\begin{equation*}
\begin{array}{l}
 W(x)=\dfrac{16\pi^2 E_{vac}}{V\Delta_1^4}=-x^2\{\ln x+y_1-\dfrac{3}{2}+\dfrac{2v}{x}+\dfrac{w}{x^2} \\[5mm]
  -4[\ln x +y_1-1+\dfrac{u}{x}]+\dfrac{x}{a_1}[\ln x +y_1-1+\dfrac{u}{x}]^2\}
\end{array}\eqno{(A.1)}
\end{equation*}
where $y_1=\ln(\Delta_1^2/\mu_s^2), a_1=\pi^2/G\Delta_1^2, u=C_2/\Delta_1^2, v=C'_2/\Delta_1^2$ and $w=C_3/\Delta_1^4$ are dimensionless constants waiting to be fixed.

The gap equation (4.6) reads
\begin{equation*}
[\ln x +y_1-1+\dfrac{u}{x}]|_{x=1}=2a,\quad\text{or}\quad u\to u_1=2a_1-y_1+1\eqno{(A.2)}
\end{equation*}
and Eq.(4.11) means that
\begin{equation*}
W(0)=0\to w=0\eqno{(A.3)}
\end{equation*}
but
\begin{equation*}
W(1)=-(y_1-\frac{3}{2}+2v_1-4a_1)<0\eqno{(A.4)}
\end{equation*}
So we need three more conditions to fix $v,y_1$ and $a_1$.

The following trick seems useful in taking derivatives, first
\begin{equation*}
\begin{array}{l}
\dfrac{dW}{dx}=W'(x)=\dfrac{2}{x}W(x)-x^2\{\dfrac{1}{x}-\dfrac{2v}{x^2}-4(\dfrac{1}{x}-\dfrac{u}{x^2})+\dfrac{1}{a}[\ln x +y_1-1+\dfrac{u}{x}]^2 \\[5mm]
+\dfrac{2x}{a_1}[\ln x +y_1-1+\dfrac{u}{x}](\dfrac{1}{x}-\dfrac{u}{x^2})\} \\[5mm]
=\dfrac{2}{x}W(x)+3x+2v-4u-\dfrac{x^2}{a}[\ln x +y_1-1+\dfrac{u}{x}]^2-\dfrac{2x}{a}(x-u)[\ln x +y_1-1+\dfrac{u}{x}]
\end{array}\eqno{(A.5)}
\end{equation*}
The condition
\begin{equation*}
W'(1)=0\eqno{(A.6)}
\end{equation*}
is imposed to get
\begin{equation*}
v\to v_1=-y_1+1+2a_1\eqno{(A.7)}
\end{equation*}
coinciding with Eq.(A.2), $v_1=u_1$. Next we have
\begin{equation*}
\begin{array}{l}
W''(x)=-\dfrac{2}{x^2}W(x)+\dfrac{2}{x}W'(x)-\dfrac{2x}{a_1}[\ln x +y_1-1+\dfrac{u}{x}]^2\\[5mm]
-\dfrac{1}{a_1}(6x-4u)[\ln x +y_1-1+\dfrac{u}{x}]-\dfrac{2}{a_1}(x-u)(1-\dfrac{u}{x})+3
\end{array}\eqno{(A.8)}
\end{equation*}
Then the condition
\begin{equation*}
W''(1)=0\eqno{(A.9)}
\end{equation*}
yields an equation
\begin{equation*}
y_1^2+a_1y_1=0\eqno{(A.10)}
\end{equation*}
so
\begin{equation*}
y_1=0\quad \text{or}\quad y_1=-a_1\eqno{(A.11)}
\end{equation*}
The further condition\footnotemark[1]\footnotetext[1]{
\begin{equation*}
\begin{array}{l}
W'''(x)=\dfrac{4}{x^3}W(x)-\dfrac{4}{x^2}W'(x)+\dfrac{2}{x}W''(x)-\dfrac{2}{a_1}[\ln x +y_1-1+\dfrac{u}{x}]^2\\[5mm]
-\dfrac{2}{a_1}[\ln x +y_1-1+\dfrac{u}{x}](5-\dfrac{2u}{x})-\dfrac{2}{a_1}(4-\dfrac{5u}{x}+\dfrac{u^2}{x^2})
\end{array}
\end{equation*}}
\begin{equation*}
W'''(1)=0\eqno{(A.12)}
\end{equation*}
yields
\begin{equation*}
y_1^2+(3-2a_1)y_1+a_1=0\eqno{(A.13)}
\end{equation*}
Combination of Eq.(A.10) with (A.13)gives the nontrivial solution
\begin{equation*}
a_1=\frac{2}{3},\quad y_1=-a_1=-\frac{2}{3}\eqno{(A.14)}
\end{equation*}
Now we can draw the whole diagram of $W(x)$ as shown in Fig.4.

The interesting problem in Fig.4 is: Besides $x_1=1$ showing the stable vacuum after VPT, there is another maximum of $W(x)$ at $x_2$ and $W(x_2)>W(x_1)$. To find $x_2$, we write Eq.(A.5)=0 explicitly (with known $u_1,v_1,y_1$ and $a_1$)
\begin{equation*}
W'(x)=x\ln x[-\dfrac{9}{2}x\ln x+12(x-1)]-\dfrac{15}{2}(x-1)^2=0\eqno{(A.15)}
\end{equation*}
Denoting $z=x\ln x$, we find
\begin{equation*}
z^2-\dfrac{8}{3}(x-1)z+\dfrac{5}{3}(x-1)^2=0\eqno{(A.16)}
\end{equation*}
which yields two solutions
\begin{equation*}
z_+=\dfrac{5}{3}(x-1),\,z_-=x-1\eqno{(A.17)}
\end{equation*}
While $z_-$ gives $x=x_1=1$ as expected, $z_+$ does give
\begin{equation*}
x\ln x=\dfrac{5}{3}(x-1)\to x=x_2=3.0841\eqno{(A.18)}
\end{equation*}
which in turn yields
\begin{equation*}
W(x_2)=-0.3842244>W(1)=-\dfrac{7}{6}=-1.1666667\eqno{(A.19)}
\end{equation*}
and
\begin{equation*}
\left\{
  \begin{array}{ll}
    W'(x_2)=1.1892\times10^{-5}\simeq0\\[2mm]
    W''(x_2)=-2.873501<0\\[2mm]
    W'''(x_2)=-6.685856<0\\[2mm]
    W''''(x_2)=-6.284411475<0
  \end{array}
\right.
\eqno{(A.20)}
\end{equation*}
as expected in contrast to
\begin{equation*}
\left\{
  \begin{array}{ll}
    W'(1)=W''(1)=W'''(1)=0\\[2mm]
    W''''(1)=9>0
  \end{array}
\right.
\eqno{(A.21)}
\end{equation*}

\section*{Appendix B: The $C$ (charge) Parity of Neutral Bosons with Fermion-Antifermion pairs}

If a neutral boson is composed of one pair of fermion-antifermion ($f\bar{f}$), \eg, the para-positronium (p-Ps) or ortho-positronium (o-Ps), what is its $C$ (charge)-parity? \footnotemark[1]\footnotetext[1]{See also from the Wikipedia of Google search: "$C$ parity" (2014). Putting neutrinos aside, we believe fermions got their masses via a process of VPT, which respects the conservation laws of $P, C, CP$ and $CPT$ individually.}

According to the definition of $C$ transformation (see, \eg, Eq.(4.129) in Ref.\cite{21}), one has
\begin{equation*}
\hat{U}(C)\hat{a}_{{\bf p}h}^\dag \hat{U}^{-1}(C)=\hat{b}_{{\bf p}h}^\dag,\quad \hat{U}(C)\hat{b}_{{\bf p}h}^\dag \hat{U}^{-1}(C)=\hat{a}_{{\bf p}h}^\dag\eqno{(B.1)}
\end{equation*}
with both the momentum $\bf p$ and helicity $h$ being unchanged. So, for one flavor case, if considering the
\begin{equation*}
 WF=\langle{\bf x}_1,{\bf x}_2,s_1,s_2,L,S,J|f\bar{f}\rangle\eqno{(B.2)}
\end{equation*}
one has
\begin{equation*}
\hat{U}(C)|f\bar{f}\rangle=(-1)^L(-1)^{S+1}(-1)|f\bar{f}\rangle=(-1)^{L+S}|f\bar{f}\rangle\eqno{(B.3)}
\end{equation*}
Here, $(-1)^L$ is due to exchange the positions of $f$ and $\bar{f}$ in space, just like that happens in a space inversion ($P$) operation:
\begin{equation*}
\hat{U}(P)|f\bar{f}\rangle=-(-1)^L|f\bar{f}\rangle\eqno{(B.4)}
\end{equation*}
with extra $(-1)$ stemming from the opposite "intrinsic parity" for $\bar{f}$ versus $f$, where $L$ is the quantum number of orbital angular momentum for the bound state $|f\bar{f}\rangle$.

The second factor $(-1)^{S+1}$ in Eq.(B.3) comes from the quantum numbers of spin-sum for $f$ and $\bar{f}$ when they are exchanged each other. The third factor $(-1)$ in Eq.(B.3) is because the operators must return back from $b^\dag a^\dag$ to $a^\dag b^\dag$. Finally, combining Eqs.(B.3) and (B.4), one has simply
\begin{equation*}
\hat{U}(C)\hat{U}(P)|f\bar{f}\rangle=(-1)^{S+1}|f\bar{f}\rangle,\quad \text{or}\quad CP=(-1)^{S+1}\eqno{(B.5)}
\end{equation*}
The bound state of $|f\bar{f}\rangle$ has a total angular momentum ${\bf J}={\bf L}+{\bf S}$ with $|L-S|\leq J\leq (L+S)$, and $J$ is often called as the "spin" of boson $|f\bar{f}\rangle$ (see p.259 in Ref.\cite{11}). For example, the ground states $(L=0)$ of positronium have
\begin{equation*}
J^{PC}=0^{-+}\quad (p-Ps,S=0)\quad \text{and}\quad J^{PC}=1^{--}\quad (o-Ps,S=1)\eqno{(B.6)}
\end{equation*}
respectively.

However, if the boson is a "collective mode" of $f\bar{f}$ configurations as shown by Eq.(6.6) or Eq.(6.8), expanding in plane-wave states, we have to check the $C$-parity of "quasiboson", Eq.(6.2) first
\begin{equation*}
\hat{U}(C)\hat{B}_{{\bf p}h}^\dag \hat{U}^{-1}(C)=\hat{B}_{{-\bf p}h}^\dag \eqno{(B.7)}
\end{equation*}
Note that while the momentum ${\bf p}\to -{\bf p}$, the helicity $h$ remains unchanged and $S=0$. Since no factor $(-1)^L$ is involved in Eq.(B.3), the quantum number $C$ reads: $C=(-1)(-1)^{S+1}=1$. Hence we have from Eqs.(6.6) and (6.8) that
\begin{equation*}
\hat{U}(C)\hat{Q}_n^{(s)\dag} \hat{U}^{-1}(C)=\hat{Q}_n^{(s)\dag},\quad \text{with}\quad J^{PC}=0^{++} \eqno{(B.8)}
\end{equation*}
and
\begin{equation*}
\hat{U}(C)\hat{Q}_n^{(ps)\dag} \hat{U}^{-1}(C)=\hat{Q}_n^{(ps)\dag},\quad \text{with}\quad J^{PC}=0^{-+} \eqno{(B.9)}
\end{equation*}
respectively.

On the other hand, for pseudovector (and vector) boson discussed in section VII, we first check from Eq.(7.1) that
\begin{equation*}
\hat{U}(C)\hat{\cal B}_{{\bf p}h}^\dag \hat{U}^{-1}(C)=\hat{\cal B}_{{-\bf p}-h}^\dag \eqno{(B.10)}
\end{equation*}
where the helicity $h\to -h$ because it is opposite in Eq.(7.1) and $S=1$, so $(-1)(-1)^{S+1}=-1$. Accordingly, we have
\begin{equation*}
\hat{U}(C)Q_n^{(pv)} \hat{U}^{-1}(C)=Q_n^{(pv)},\quad (J^{PC}=1^{++}) \eqno{(B.11)}
\end{equation*}
\begin{equation*}
\hat{U}(C)Q_n^{(v)} \hat{U}^{-1}(C)=-Q_n^{(v)},\quad (J^{PC}=1^{--}) \eqno{(B.12)}
\end{equation*}
In summary, we find four possible collective modes for neutral bosons in this paper. They have $J^{PC}=0^{++},0^{-+},1^{++}$ and $1^{--}$ respectively. For lepton case, it seems that only $J^{PC}=0^{++}$ with $\Omega_n^{(s)}=2\Delta_1$, Eq.(6.17), is a massive boson whereas other three are Goldstone bosons. But for quark case, all of these collective modes have to be considered as candidates for massive neutral mesons.

\section*{Appendix C: Calculations of the Signature for Chiral Symmetry Breaking}

For proving the signature for VPT, Eqs.(8.6)-(8.9), we need the WFs in the field operators, Eqs.(8.1)-(8.5). According to discussions on the Dirac equation in textbooks like \cite{17,21} and Ref.\cite{20}, for a free fermion, either massless or massive, the 4-component spinor, $u_h({\bf p}) [v_h({\bf p})]$ attributed to $e^{i({\bf p}\cdot{\bf x}-E_pt)} [e^{-i({\bf p}\cdot{\bf x}-E_pt)}]$ referring to a fermion (antifermion) can be chosen as the eigenstate of helicity $h$ generally
\begin{equation*}
u_h({\bf p})=N\left(
                \begin{array}{c}
                  \xi_h({\bf p}) \\
                  \frac{hp}{m+E_p}\xi_h({\bf p}) \\
                \end{array}
              \right),\quad v_h({\bf p})=\gamma_5u_{-h}({\bf p})=N\left(
                \begin{array}{c}
                 \frac{-hp}{m+E_p}\xi_{-h}({\bf p}) \\
                  \xi_{-h}({\bf p}) \\
                \end{array}
              \right)\eqno{(C.1)}
\end{equation*}
where $N=\sqrt{\frac{E_p+m}{2E_p}}, h=\frac{{\boldsymbol\sigma}\cdot{\bf p}}{|{\bf p}|}=\frac{{\boldsymbol\sigma}\cdot{\bf p}}{p}$. Notice that because ${\boldsymbol\sigma}_c=-{\boldsymbol\sigma}$ as proved in Ref.\cite{20}, the helicity in $v$ is just opposite to that in $u$ as shown by Eq.(C.1) (for clarity, the subscript "c" is omitted for $h,{\bf p}$ and $E_p=\sqrt{p^2+m^2}$ in the WF of antifermion. see Eqs.(6.12)-(6.17) in \cite{20}). And the 2-component spinor $\xi_h({\bf p})$ is chosen as
\begin{equation*}
\xi_{h=1}=\left(
                \begin{array}{c}
                  \cos\frac{\theta}{2} \\
                  \sin\frac{\theta}{2}e^{i\phi} \\
                \end{array}
              \right),\quad \xi_{h=-1}=\left(
                \begin{array}{c}
                  \sin\frac{\theta}{2} \\
                  -\cos\frac{\theta}{2}e^{i\phi} \\
                \end{array}
              \right)=\xi_h({-\bf p})\eqno{(C.2)}
\end{equation*}
where ${\bf p}(p,\theta,\phi)$ with $(\theta,\phi)$ being the angles of ${\bf p}$ in spherical coordinates, and
\begin{equation*}
 \xi_h({\bf p})^\dag\xi_{h'}({\bf p})=\delta_{hh'},\quad ({\bf p}\ \text{not summed})\eqno{(C.3)}
\end{equation*}
In Eq.(C.1), if the mass $m=\Delta_1\neq0$, the normalization coefficient $N$ is chosen such that
\begin{equation*}
 u_h({\bf p})^\dag u_{h'}({\bf p})=v_h({\bf p})^\dag v_{h'}({\bf p})=\delta_{hh'},\quad ({\bf p}\ \text{not summed})\eqno{(C.4)}
\end{equation*}
When $m=0$ before the VPT, Eqs.(C.1)-(C.4) remain valid by just setting $m=0$. Another useful formulas are
\begin{equation*}
 \bar{u}_h({-\bf p}) = \gamma^0u_{-h}({\bf p}),\quad \bar{v}_h({-\bf p}) =- \gamma^0v_{-h}({\bf p})\eqno{(C.5)}
\end{equation*}
\begin{equation*}
  \bar{u}_h({\bf p}) u_{h'}({\bf p}) = \bar{v}_h({\bf p}) v_{h'}({\bf p})=0 \eqno{(C.6)}
\end{equation*}
\begin{equation*}
  \bar{u}_h({\bf p}) u_{h'}({-\bf p})  =\delta_{h,-h'},\quad  \bar{v}_h({\bf p}) v_{h'}({-\bf p})  =-\delta_{h,-h'}\eqno{(C.7)}
\end{equation*}
\begin{equation*}
 \bar{u}_h({\bf p}) v_{h'}({\bf p}) = \bar{v}_h({\bf p}) u_{h'}({\bf p})=0 \eqno{(C.8)}
\end{equation*}
\begin{equation*}
  \bar{u}_h({\bf p}) v_{h'}({-\bf p}) =\bar{u}_h({\bf p}) v_{h'}({-\bf p})=-h\delta_{hh'} \eqno{(C.9)}
\end{equation*}
\begin{equation*}
   \bar{u}_h({\bf p})\gamma_5 u_{h'}({\bf p})  = \bar{u}_h({\bf p})\gamma_5 u_{h'}({\bf p}) =0\eqno{(C.10)}
\end{equation*}
\begin{equation*}
 \bar{u}_h({\bf p})\gamma_5 v_{h'}({\bf p})  = \bar{v}_h({\bf p})\gamma_5 u_{h'}({\bf p})=0  \eqno{(C.11)}
\end{equation*}
\begin{equation*}
  \bar{u}_h({\bf p})\gamma_5 v_{h'}({-\bf p})  =\delta_{hh'},\quad  \bar{v}_h({\bf p})\gamma_5 u_{h'}({-\bf p})=-\delta_{hh'}\eqno{(C.12)}
\end{equation*}
Then for $m=0$ case, substituting Eq.(2.1) and using $\int d{\bf x}e^{i({\bf p}+{\bf p}')\cdot{\bf x}}=V\delta_{\bf p,-p'}$, we have
\begin{equation*}
\int d{\bf x}:\hat{\bar\psi}(x)\hat{\psi}(x):=-\sum_{{\bf p},h}h(\hat{a}_{{\bf p}h}^\dag \hat{b}_{{-\bf p}h}^\dag+\hat{b}_{{-\bf p}h}\hat{a}_{{\bf p}h}) \eqno{(C.13)}
\end{equation*}
where Eqs.(C.6) and (C.9) had been used. So Eq.(8.7) follows immediately.

After VPT, we can prove Eq.(8.8) since
\begin{equation*}
\langle\tilde{0}|\int d{\bf x}:\hat{\bar\psi}(x)\hat{\psi}(x):|\tilde0\rangle=-4\sum_{\bf p}U_pV_p=-2\sum_{\bf p}\dfrac{\Delta_1}{E_p}=-\dfrac{\Delta_1 V}{2G}\eqno{(C.14)}
\end{equation*}
due to Eqs.(3.7), (3.22) and (3.23). Eventually, the signature after VPT reads
\begin{equation*}
\langle\tilde{0}|{\bar\psi}\psi|\tilde0\rangle\equiv\dfrac{1}{V}\langle\tilde{0}|\int d{\bf x}:\hat{\bar\psi}(x)\hat{\psi}(x):|\tilde0\rangle=-\dfrac{\Delta_1^3 }{3\pi^2}\eqno{(C.15)}
\end{equation*}
where Eq.(4.35) had been used. After VPT, instead of Eqs.(C.13)-(C.14), we may use Eq.(8.5) yielding
\begin{equation*}
\langle\tilde{0}|\int d{\bf x}:\hat{\bar\Psi}(x)\hat{\Psi}(x):|\tilde0\rangle=-\langle\tilde{0}|\sum_{{\bf p},h}h(\hat{\alpha}_{{\bf p}h}^\dag \hat{\beta}_{{-\bf p}h}^\dag+\hat{\beta}_{{-\bf p}h}\hat{\alpha}_{{\bf p}h})|\tilde0\rangle=-\dfrac{\Delta_1V}{2G}\eqno{(C.16)}
\end{equation*}
where Eq.(5.1) had been used. Then we find ($''~\hat{}~''$ will be erased)
\begin{equation*}
\langle\tilde{0}|{\bar\Psi}\Psi|\tilde0\rangle=-\dfrac{\Delta_1^3 }{3\pi^2}\eqno{(C.17)}
\end{equation*}
coinciding with Eq.(C.14) as shown by Eq.(8.8). Similarly, because of
\begin{equation*}
\int d{\bf x}{\bar\psi}(x)\gamma_5\psi(x)=\sum_{{\bf p},h}(\hat{a}_{{\bf p}h}^\dag \hat{b}_{{-\bf p}h}^\dag-\hat{b}_{{-\bf p}h}\hat{a}_{{\bf p}h}) \eqno{(C.18)}
\end{equation*}
\begin{equation*}
\int d{\bf x}{\bar\Psi}(x)\gamma_5\Psi(x)=\sum_{{\bf p},h}(\hat{\alpha}_{{\bf p}h}^\dag \hat{\beta}_{{-\bf p}h}^\dag-\hat{\beta}_{{-\bf p}h}\hat{\alpha}_{{\bf p}h}) \eqno{(C.19)}
\end{equation*}
and Eq.(3.7), Eq.(8.9) are proved.

\vspace*{-2mm}

\section*{Acknowledgements}

\vspace*{-2mm}

Being a little late, we wish to dedicate this paper to Y. Nambu (1921-2015) for his idea has been inspiring us in the past decades.

We also thank J. Abramson, E. Bodegom, S. Q. Chen, Y. X. Chen, X. X. Dai, V. Dvoeglazov, S. S. Feng, J. Freeouf, R. T. Fu, F. Han, J. Jiao, T. C. Kerrigan, A. Khalil, R. Koenenkamp, D. X. Kong, A. La Rosa, J. S. Leung, P. T. Leung, H. Z. Li, Q. G. Lin, W. F. Lu, D. Mitchell, J. H. Ruan, E. Sanchez, Z. Y. Shen, Z. Q. Shi, P. Smejtek, X. T. Song, R. K. Su, X. Sun, R. B. Tao, H. B. Wang, Y. S. Wang, X. Xue, F. J. Yang and G. H. Yang for encouragement, collaborations and helpful discussions.


\newpage

\begin{figure}
  \centering
  \includegraphics{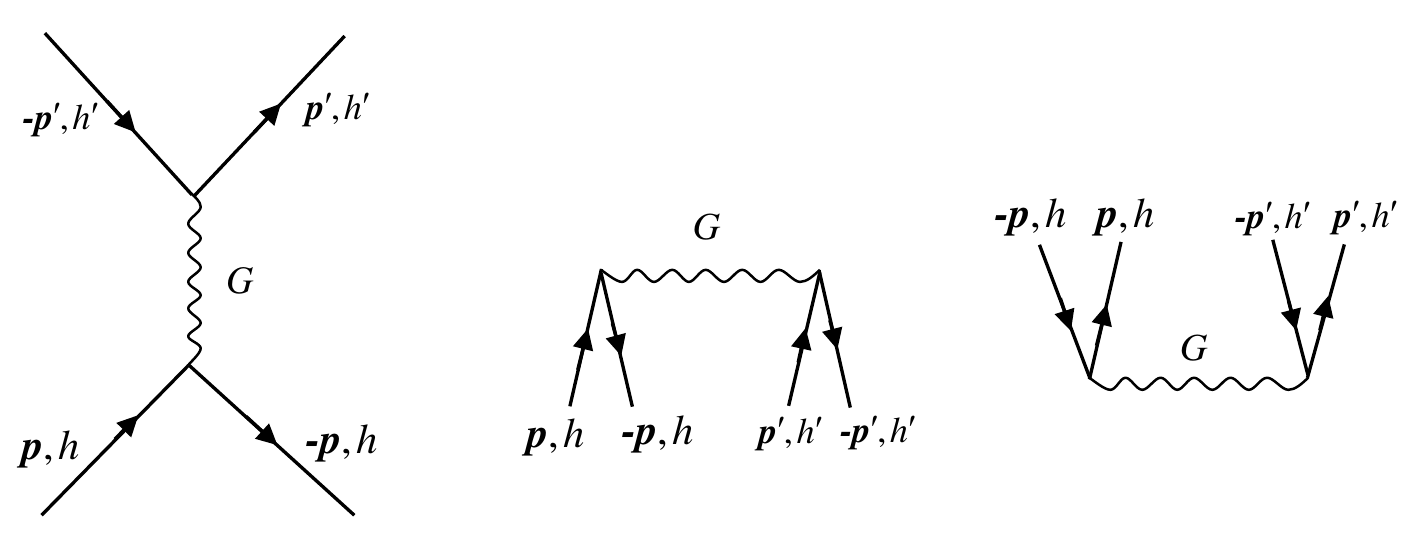}\\
  \caption{Three types of Feynman diagram for the effective Hamiltonian, Eq.(2.7), responsible for the VPT, with $G\sim [M]^{-2}$ for lepton. }\label{fig1}
\end{figure}

\begin{figure}
  \centering
  \includegraphics{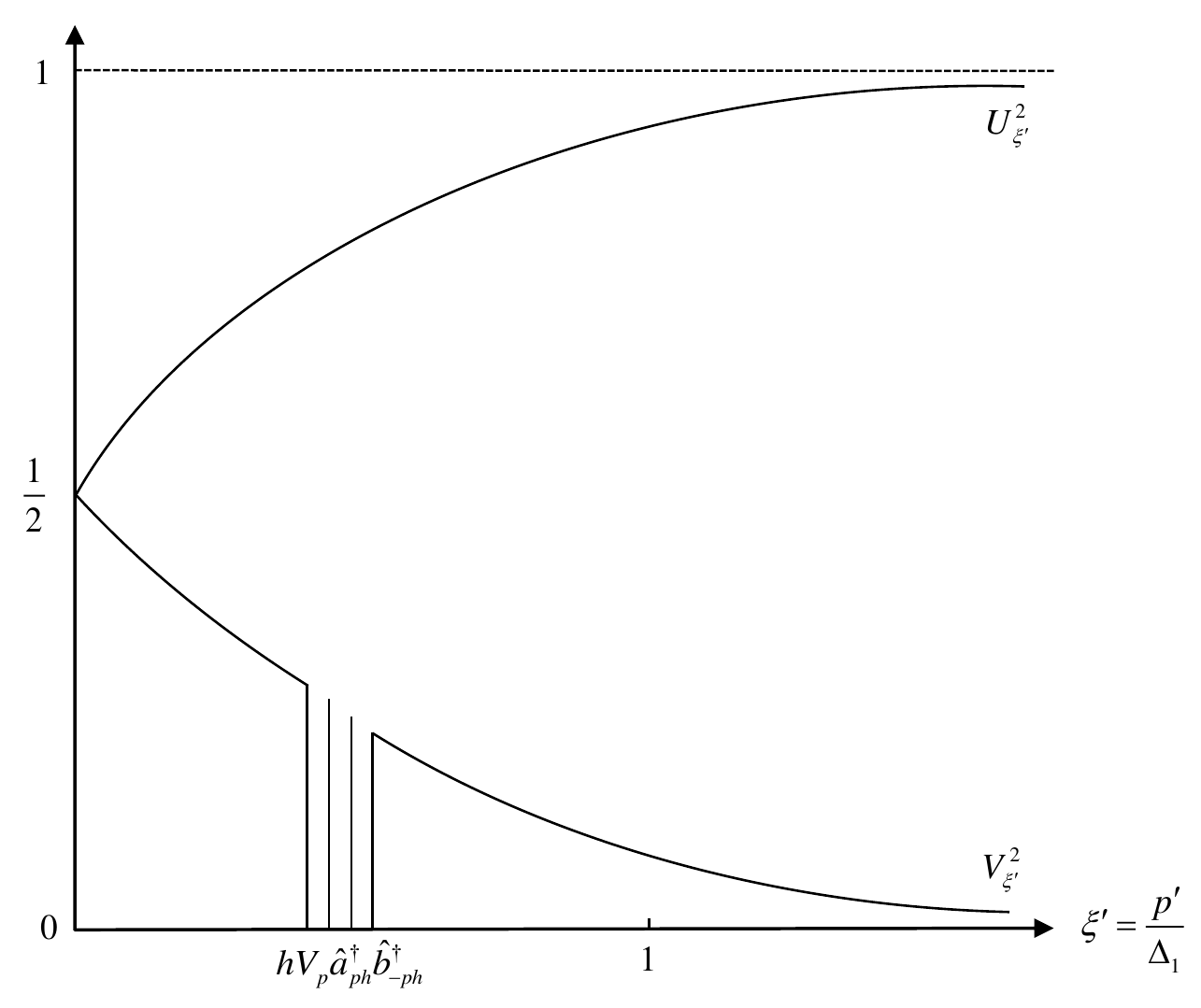}\\
  \caption{$V_p^2=V_\xi^2=\frac{1}{2}(1-\frac{\xi}{\sqrt{1+\xi^2}})$ and $U_p^2=U_\xi^2=\frac{1}{2}(1+\frac{\xi}{\sqrt{1+\xi^2}})$ as functions of $\xi=\frac{p}{\Delta_1}$ in the NJLVS (helicity $h$ is not shown for clarity).}\label{fig2}
\end{figure}

\begin{figure}
  \centering
  \includegraphics{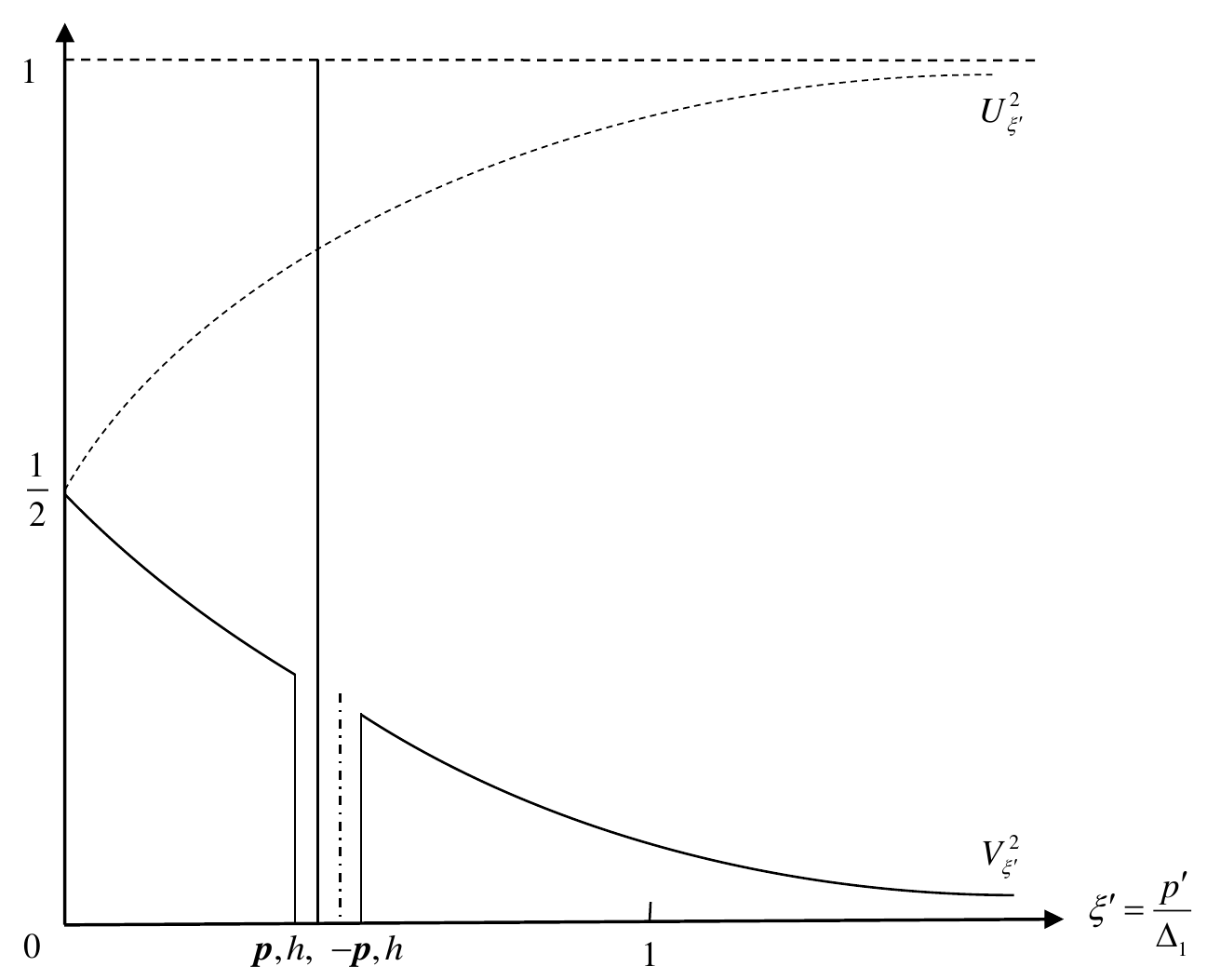}\\
  \caption{The "physical meaning" of the creation operator for the massive quasi-particle $\hat{\alpha}^\dag_{{\bf p}h}=U_p\hat{a}^\dag_{{\bf p}h}-hV_p\hat{b}_{{-\bf p}h}$ in NJLT, Eq.(5.11), is shown on the background of NJLVS (helicity $h$ is not shown for clarity). (see Fig.2)}\label{fig3}
\end{figure}

\begin{figure}
  \centering
  \includegraphics{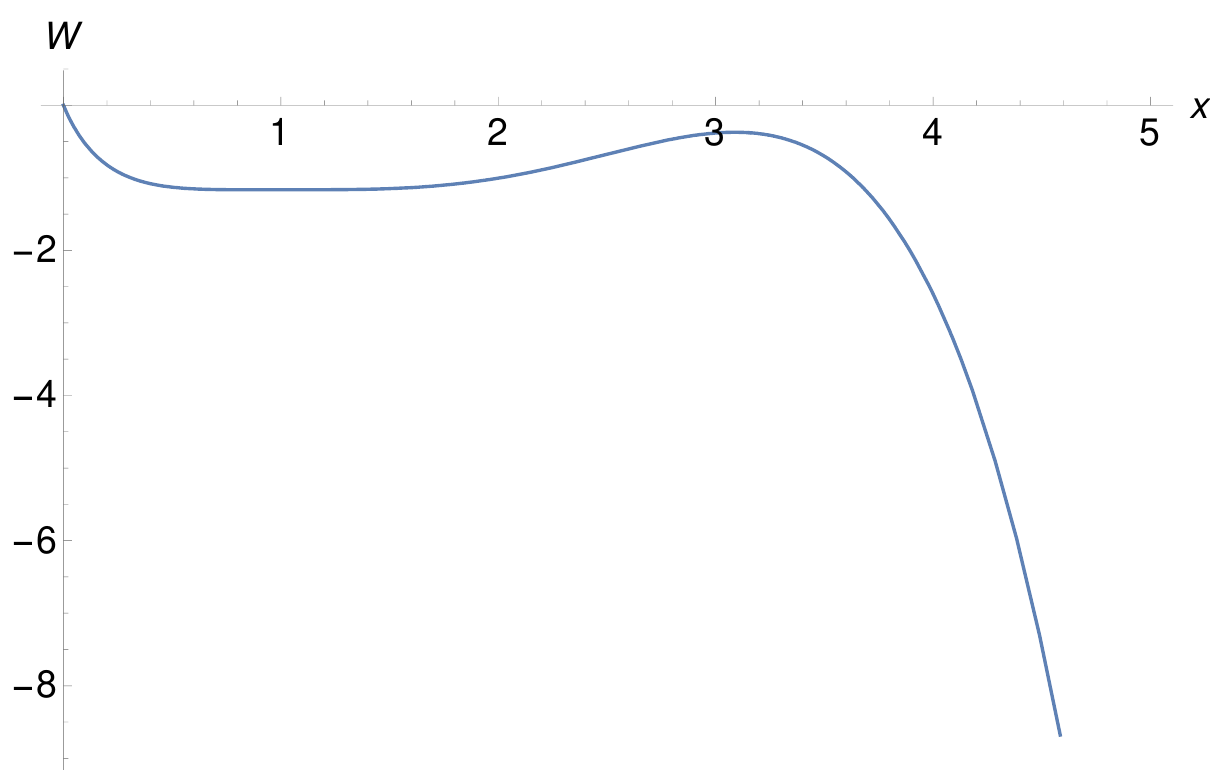}\\
  \caption{The dimensionless energy density of vacuum  $W(x)=3x^2\ln x+6x-\frac{9}{2}x^2-\frac{3}{2}x[x(\ln x-\frac{5}{3})+3]^2$ as a function of $x=(\frac{\Delta}{\Delta_1})^2$ in NJL model, see Eq.(A.1).}\label{fig4}
\end{figure}

\end{document}